\documentclass[a4paper,11pt]{article}
\pdfoutput=1 % if your are submitting a pdflatex (i.e. if you have
\usepackage{jheppub} % for details on the use of the package, please
\usepackage[T1]{fontenc} % if needed
\RequirePackage[numbers,sort&compress]{natbib}
\usepackage{cleveref}
\usepackage{float}
\usepackage{amsmath}
\usepackage{amssymb}
\usepackage{mathtools}
\usepackage{siunitx}
\usepackage{amsfonts}
\usepackage{dsfont}
\usepackage{array}
\usepackage{bm}
\usepackage{mathrsfs}
\usepackage{pifont}
\usepackage{multirow}
\usepackage{upgreek}
\usepackage{verbatim}
\usepackage{slashed}
\usepackage{ucs}
\usepackage{subfigure}
\usepackage{afterpage}
\usepackage[dvipsnames]{xcolor}
\definecolor{mygreen}{rgb}{0, 0.4, 0}
\graphicspath{{./pics/}}
\newcommand{\nn}{\nonumber}

\renewcommand{\Re}{{\rm Re}\,}
\renewcommand{\Im}{{\rm Im}\,}

\newcommand{\cM}[0]{\mathcal M}

\newcommand{\cK}[0]{\mathcal K}
\newcommand{\cD}[0]{\mathcal D}
\newcommand{\cS}[0]{\mathcal S}

\newcommand{\cL}[0]{\mathcal L}
\newcommand{\cR}[0]{\mathcal R}

\newcommand{\wt}[0]{\widetilde}
\newcommand{\PV}[0]{{\rm PV}}
\newcommand{\iso}[0]{{\rm iso}}
\newcommand{\df}[0]{{\rm df}}

\newcommand{\Kiso}[0]{{\cK_{\df,3}^{\iso}}}

\newcommand{\Mdf}[0]{{\cM_{\df,3}}}

\newcommand{\Kdf}[0]{{\cK_{\df,3}}}

\makeatletter
\renewcommand*\env@matrix[1][\arraystretch]{%
  \edef\arraystretch{#1}%
  \hskip -\arraycolsep
  \let\@ifnextchar\new@ifnextchar
  \array{*\c@MaxMatrixCols c}}
\makeatother

\title{\boldmath Three-body resonances in the $\varphi^4$ theory}
\author[a,1]{Marco Garofalo
}
\author[a,b]{, Maxim Mai}
\author[c]{, Fernando Romero-L\'opez}
\author[a,d]{, Akaki Rusetsky}
\author[a]{ and Carsten Urbach}

\affiliation[a]{Helmholtz-Institut f\"ur Strahlen- und Kernphysik (Theorie) and Bethe Center for Theoretical Physics, Universit\"at Bonn, D--53115 Bonn, Germany}
\affiliation[c]{Center for Theoretical Physics, Massachusetts Institute of Technology, Cambridge, MA 02139, USA}
\affiliation[b]{Institute for Nuclear Studies and Department of Physics, The George Washington University, Washington, DC 20052, USA}
\affiliation[d]{Tbilisi State University, 0186 Tbilisi, Georgia}

\emailAdd{garofalo@hiskp.uni-bonn.de}
\abstract{We study the properties of three-body resonances using a lattice complex scalar $\varphi^4$ theory with two scalars, with parameters chosen such that one heavy particle can decay into three light ones. We determine the two- and three-body spectra for several lattice volumes using variational techniques, and then analyze them with two versions of the three-particle finite-volume formalism: the Relativistic Field Theory approach and the Finite-Volume Unitarity approach. We find that both methods provide an equivalent description of the energy levels, and we are able to fit the spectra using simple parametrizations of the scattering quantities. By solving the integral equations of the corresponding three-particle formalisms, we determine the pole position of the resonance in the complex energy plane and thereby its mass and width. We find very good agreement between the two methods at different values of the coupling of the theory.}
\preprint{MIT-CTP/5487}

\begin{document} 
\maketitle
\flushbottom
%%%%%%%%%%%%%%%%%%%%%%%%%%%%%%%%%%%%%%%%%%%%%%%%%%%%

%%%%%%%%%%%%%%%%%%%%%%%%%%%%%%%%%%%%%%%%%%%%%%%%%%

%%%%%%%%%%%%%%%%%%%%%%%%%%%%%%%%%%%%%%%%%%%%%%%%%%
\section{Introduction}
\label{sec:intro}
%%%%%%%%%%%%%%%%%%%%%%%%%%%%%%%%%%%%%%%%%%%%%%%%%%

The hadronic three-body problem marks the current frontier of the theoretical and computational progress in hadron spectroscopy. Its understanding is crucial to various systems of high relevance, such as the Roper resonance and its large branching ratios to $\pi\pi N$ channels~\cite{Arndt:2006bf, Ceci:2011ae, Loring:2001kx, Lang:2016hnn, Severt:2022jtg}
and other mesonic resonances, such
as the $\omega(782)$  \cite{10.1093/ptep/ptac097} decaying into $\pi^+\pi^-\pi^0$,
or the search for spin exotics~\cite{Dobbs:2019sgr} decaying only to three-body final states.

%just to name a few.

Understanding such systems 
%from first principles
has been a major challenge for a long time, and only recently it came into reach due to rapid theoretical and computational advances. Specifically, enormous progress has been achieved connecting the spectrum of three-body systems in finite and infinite volume~\cite{Beane:2007es,Polejaeva:2012ut, Briceno:2012rv, Roca:2012rx, Bour:2012hn, Meissner:2014dea, Jansen:2015lha, Hansen:2014eka, Hansen:2015zga, Hansen:2015zta, Hansen:2016fzj, Guo:2016fgl, Konig:2017krd, Hammer:2017uqm, Hammer:2017kms, Briceno:2017tce, Sharpe:2017jej, Guo:2017crd, Guo:2017ism, Meng:2017jgx, Guo:2018ibd, Guo:2018xbv, Klos:2018sen, Briceno:2018mlh, Briceno:2018aml, Mai:2017bge, Mai:2018djl, Doring:2018xxx, Jackura:2019bmu, Mai:2019fba, Guo:2019hih, Blanton:2019igq, Briceno:2019muc, Romero-Lopez:2019qrt, Pang:2019dfe, Guo:2019ogp, Zhu:2019dho, Pang:2020pkl, Hansen:2020zhy, Guo:2020spn, Guo:2020wbl, Guo:2020ikh, Guo:2020kph, Blanton:2020gha, Blanton:2020gmf, Muller:2020vtt, Brett:2021wyd,Muller:2021uur, Grabowska:2021xkp, Muller:2020wjo,Jackura:2020bsk, Hansen:2021ofl, Blanton:2021mih,Blanton:2021eyf,Jackura:2022gib} via the so-called \emph{quantization condition}, for reviews see Refs.~\cite{Hansen:2019nir, Rusetsky:2019gyk,Mai:2021lwb,Romero-Lopez:2021zdo,Mai:2022eur,Romero-Lopez:2022usb}.
Several weakly interacting systems of three mesons at maximal isospin have indeed been calculated in Lattice QCD~\cite{Mai:2018djl, Horz:2019rrn, Blanton:2019vdk, Mai:2019fba, Culver:2019vvu, Fischer:2020jzp, Hansen:2020otl, Alexandru:2020xqf, Brett:2021wyd, Blanton:2021llb, NPLQCD:2020ozd} and successfully studied using three-body quantization conditions. The strategy for resonant three-body systems has been laid out~\cite{Hansen:2020zhy, Mai:2021nul, Severt:2022eic}, and its first application to the axial $a_1(1260)$-resonance from Lattice QCD has been accomplished~\cite{Mai:2021nul}. 

With only a few available Lattice QCD results for resonant three-body systems~\cite{Lang:2016hnn, Mai:2021nul}, it was so far not possible to uncover the appearance of such interesting effects as the avoided level crossing or test various strategies in extracting infinite-volume quantities. In this work, we attempt to close this gap by using $\varphi^4$-theory, which is a viable testbed for formalism in a controlled setup. For previous work see Refs.~\cite{Romero-Lopez:2018rcb, Romero-Lopez:2020rdq, Garofalo:2021bzl}. 
Due to reduced computational costs compared to lattice QCD, the exploration of parameter space is much more flexible. In addition, one can control the particle content and the resonance parameters freely.

On the analysis side, we utilize state-of-the-art quantization conditions, namely the \emph{Relativistic Field Theory} (RFT)~\cite{Hansen:2014eka, Hansen:2015zga} and \emph{Finite-Volume Unitarity} (FVU)~\cite{Mai:2017bge, Mai:2018djl} approach, testing  and comparing those for the first time on the same set of data. We note that this also tests the performance of the \emph{Non-Relativistic Effective Field Theory} (NREFT)~\cite{Hammer:2017uqm, Hammer:2017kms} approach, which in its Lorentz-invariant formulation~\cite{Muller:2021uur} is algebraically identical to the FVU quantization condition, at least within the approximations used in this work. At their core, all these approaches aim to separate the power-law volume dependence of the three-particle scattering amplitude from the exponentially suppressed one, which necessitates singling out the classes of Feynman diagrams
in which the intermediate particles can go on-shell.  Obviously, this goal is achieved by different means in the different formalisms, but the final result is equivalent in the sense that all configurations of three particles being on-shell are accounted for. The differences include the chosen cutoff in the spectator momentum, how exponentially suppressed terms are accounted for, and the particular choice of the parametrization for the (sub)-system dynamics. Relations between different parametrizations are, in general, non-trivial and involve integral equations, see Refs.~\cite{Blanton:2020jnm, Brett:2021wyd, Jackura:2022gib}.  Still, an empirical comparison of both finite-volume approaches (FVU/RFT) on the same set of lattice results has not been performed. 
Potentially, there might be scenarios where one or another approach may be more advantageous to describe the actual data.
This defines the second goal of this work, allowing one to identify possible systematics of analysis tools for future studies.

This paper is organized as follows: in section~\ref{sec:phi4model} we show the toy model with an explicit resonance coupling to the three-particle final states. In section~\ref{sec:QCs} we recap both finite-volume quantization conditions and discuss the workflow for extraction of scattering parameters. Next, in section~\ref{sec:results} we present our analysis of the finite-volume spectrum at different values of the couplings in the action. Furthermore, in section~\ref{sec:infinitevolume} we present the determination of the mass and width of the resonance based on the fitted scattering quantities.
We conclude with the summary in section~\ref{sec:conclu}.

%%%%%%%%%%%%%%%%%%%%%%%%%%%%%%%%%%%%%%%%%%%%%%%%%%
\section{Description of the Model}
\label{sec:phi4model}
%%%%%%%%%%%%%%%%%%%%%%%%%%%%%%%%%%%%%%%%%%%%%%%%%%

Scalar models were already used to study resonances of two particles in \cite{GATTRINGER199295, RUMMUKAINEN1995397}, providing the necessary background for carrying out similar studies also in lattice QCD. Here we 
study a similar model which has a resonance coupled to the three-body state.
The Euclidean model is composed of two complex scalar fields $\varphi_i$ ($i=0,1$) with non-degenerate (bare) masses $m_0<m_1$ and Lagrangian 
  \begin{align}
  {\cal L}=  \sum_{i=0,1}\left[\frac{1}{2} \partial^\mu \varphi_i^\dagger\partial_\mu \varphi_i +\frac{1}{2}m_i^2 \varphi_i^\dagger\varphi_i +\lambda_i (\varphi_i^\dagger\varphi_i)^2 \right]
  +{\frac{g}{2} \varphi_1^\dagger \varphi_0^3}+ \text{h.c.}\,.
\label{eq:lagrangian} 
 \end{align}
The terms proportional to the bare coupling $g$
make the heavy particle (with field $\varphi_1$) unstable since it can decay into three light particles, each of them associated with field $\varphi_0$.
The Lagrangian has a global symmetry $\varphi_0\to e^{i\alpha}\varphi_0$ and 
$\varphi_1\to e^{i3\alpha}\varphi_1$, which prevents the mixing of operators transforming 
in different ways under this symmetry, for instance, the operator $\varphi_0$ cannot mix with the operator $\varphi_0^3$.
Thus, the mixing of one light particle with three light particles is forbidden. This will be useful for the extraction of the spectrum.

To study the problem numerically,
we define the theory on a finite hypercubic lattice with periodic boundary conditions, lattice spacing $a$, and volume $V=T \cdot L^3$, where $T$ denotes the Euclidean time extent and $L$ the spatial extent of the lattice. We define the derivatives in the Lagrangian on a lattice as finite differences 
\begin{equation}
    \partial_\mu \varphi_i(x)=\frac{1}{a}\,(\varphi_i(x+a\mu)-\varphi_i(x))\,.
\end{equation}
 In the following, we set the lattice spacing $a=1$ for convenience.
 Redefining
 \begin{equation}
     m_i^2=\frac{1-2\hat\lambda_i}{\kappa_i}-8, \quad \hat\lambda_i=4\kappa_i^2\lambda_i, \quad \hat g= 4g\sqrt{\kappa_0^3 \kappa_1} , \quad \varphi_i=\sqrt{2\kappa_i}\phi_i,
 \end{equation}
%$m_i=\frac{1-2\lambda_i}{\kappa_i}-8$, $\hat\lambda_i=\frac{\lambda_i}{4\kappa_i^2}$, $\hat g= \frac{1}{4\sqrt{\kappa_0 \kappa_1^3} }$ and $\varphi_i=\sqrt{2\kappa_i}\phi_i$. 
the discretized lattice action reads
\begin{align}
\begin{split}
  S=\sum_x  \Bigg \{ \sum_{i=0,1}\bigg[  
  &-\kappa_i\phi_i^\dagger(x)\sum_\mu  \left[ \phi_i(x+\mu)+ \phi_i(x-\mu)\right] + \phi_i^\dagger(x)\phi_i(x) + \hat\lambda_i (\phi_i^\dagger(x) \phi_i(x)-1)^2 \bigg] \\ &+ \frac{\hat g}{2} \phi_1^\dagger(x) \phi_0^3(x)+ \text{h.c. }  \Bigg \}.
  \end{split}
 \end{align}
As a further simplification, we study the model in the limit $\lambda_i\to \infty$
for both couplings $i=0,1$, this is often referred to as the Ising limit (a more detailed discussion of this limit can be found in section 2.4.2 of \cite{montvay1994quantum}). In this limit, the only non-zero contribution to the path integral over the field comes from the configurations that satisfy $\phi^\dagger_i(x)\phi_i(x)=1$. Thus, the integral over the complex variable $\phi_i$ is reduced to the integral of an angle $\theta$, representing the phase of the field $\phi_i=e^{i\theta_i}$ and the action simplifies to 
\begin{align}\label{eq:code-action}
  S=\sum_x  \Bigg \{ \sum_{i=0,1}\bigg[  
  -\kappa_i\phi_i^\dagger(x)\sum_\mu  \left[ \phi_i(x+\mu)+ \phi_i(x-\mu)\right] \bigg]+ \frac{\hat g}{2} \phi_1^\dagger(x) \phi_0^3(x)+ \text{h.c. }  \Bigg \}.
 \end{align} 

The model is most likely trivial and reduces to a free theory in the continuum limit \cite{LUSCHER198725}. However, with a small but finite lattice spacing and with energies below the cutoff scale, the model effectively describes an interacting continuum field theory.

\subsection{Simulation algorithm}
We generate ensembles, using the Metropolis-Hastings
algorithm\footnote{For this model, more advanced algorithms are available, see e.g. Ref.~\cite{Wolff:1988uh}. However, given the large values of the bare mass in our ensembles (Table~\ref{tab:ensembles}), we do not expect a significant speed-up compared to the Metropolis-Hastings algorithm.
Our implementation is available at \url{https://github.com/HISKP-LQCD/Z2-phi4/tree/complex-ising}}. For each point $x$ a new configuration is proposed $\phi'(x)$ from a uniform distribution, and it is accepted with probability 
$P=\min\left\{1, \exp\left(- \Delta S\right)\right\}$, where
$\Delta S$ is the variation of the action (\ref{eq:code-action}).
When simulating even lattice sizes, 
the lattice can be divided, as usual, into two sectors (even/odd), where all the points within one sector can be updated in parallel. This strategy cannot be used for odd lattice sizes, where we have to divide the lattice into three sectors instead, which can be updated simultaneously. 
We implement the simulation algorithm using the Kokkos C++ library~\cite{9485033, 9502936}, which provides abstractions  for both parallel execution of code and data management in order to write performance-portable applications.
The list of ensembles, generated in this work, 
is compiled in Table~\ref{tab:ensembles}.

We used $2\cdot10^{7}$ configurations for each ensemble, generated from 200 replicas each of $10^5$ thermalized configurations.
We skip the first $10^4$ configurations in each replica for thermalization.
For the light mass $M_0$, we measured the integrated autocorrelation time  $\tau_\mathrm{int}\sim0.8$, in units of the Monte Carlo time.
We bin the configurations in blocks of $10^5$ (the entire replica), which we expect to be safely larger than the $\tau_\mathrm{int}$ of any of the observables investigated here.
We resample the resulting 200 configurations using the Jackknife 
technique to propagate uncertainties to all derived quantities.

\begin{table}[t]
    \centering
    \begin{tabular}{cccc}
    \hline
             $\kappa_0$ & $\kappa_1$ & $g$ & $L$ \\
             \hline
             0.148522   & 0.134228   &  0     &  20-24\\
             0.147957   & 0.131234   &  4.43  &  20-25\\
             0.147710   & 0.131062   &  8.87  &  21-26 \\
             0.147145   & 0.131062   &  17.81 &  21-27\\  \hline
    \end{tabular}
    \caption{Ensembles used in this work. The time extent is always kept fixed to $T=64$.  }
    \label{tab:ensembles}
\end{table}

%%%%%%%%%%%%%%%%%%%%%%%%%%%%%%
\subsection{Observables}
%%%%%%%%%%%%%%%%%%%%%%%%%%%%%%

We measure the mass of the light particle from an exponential fit to the appropriate two-point correlation functions at large time distances 
\begin{equation}
\langle \tilde\phi_{0}^\dagger(t)\tilde\phi_{0}(0) \rangle\ \approx\
   |A_{\phi_0\to 0}|^2 \left(e^{ - M_{0}  t}  + e^{ - M_{0} (T - t)}\right)\,,    
\end{equation}
with $\tilde \phi_{i}(t)=\sum_{\bm x}\phi_{i}(t,\bm x)$ being a field projected onto the zero spatial momentum, $M_0$ the mass of one particle $\phi_0$ and the matrix element 
$A_{\phi_0\to 0}=\langle \phi_0| \tilde\phi_0^\dagger| 0\rangle$.   
Analogously, the energy of the two light particle system $E_2$ can be determined from
\begin{align}
     \langle \big[\tilde\phi_0^\dagger(t)\big]^2\big[\tilde\phi_0(0)\big]^2 \rangle\approx |A_{2\phi_0\to0}|^2\left(e^{-E_2t}-e^{-E_2(T-t)}\right)
        + |A_{\phi_0\to\phi_0}|^2 e^{-M_0 T}\,\,
\end{align}
where the matrix elements are $A_{2\phi_0\to0}=\langle 2\phi_0| \big[\tilde\phi_0^\dagger(t)\big]^2| 0\rangle$ and $A_{\phi_0\to\phi_0}=\langle \phi_0| \big[\tilde\phi_0^\dagger(t)\big]^2| \phi_0^\dagger\rangle$.
When the coupling is zero $g=0$ the particle $\phi_1$ is stable thus its mass $M_1$ can be measured as for $M_0$ from the exponential fit of
\begin{equation}\label{eq:M1}
\langle \tilde\phi_{1}^\dagger(t)\tilde\phi_{1}(0) \rangle \approx
   A_{1\to 0} \left(e^{ - M_{1}  t}  + e^{ - M_{1} (T - t)}\right)\,.
\end{equation}
If $g>0$, then one $\phi_1$ particle can mix with the three $\phi_0$ particles. Hence we consider the operators $\phi_0^3$ and $\phi_1$ with identical quantum numbers to construct the correlator matrix
\begin{equation}
        {\cal C}(t)=\begin{pmatrix}[1.2] % in case you need more height
        \langle\big[ \tilde\phi_0^\dagger(t)\big]^3\big[\tilde\phi_0(0)\big]^3 \rangle & 
        \langle\big[ \tilde\phi_0^\dagger(t)\big]^3\tilde\phi_1(0) \rangle\\
         \langle \tilde\phi_1^\dagger(t)\big[\tilde\phi_0(0)\big]^3 \rangle & \langle \tilde\phi_1^\dagger(t)\tilde\phi_1(0) \rangle
        \end{pmatrix} \,.
\end{equation}
We solve the generalized eigenvalue problem (GEVP)~\cite{Michael:1982gb,Luscher:1990ck,GEVP,Fischer:2020bgv}% given by
\begin{equation}
        {\cal C}(t)\, v_n=\lambda(t,t_0)\, {\cal C}(t_0)v_n\,
\end{equation}
for all $t$, keeping fixed $t_0=3$ in lattice units. From the eigenvalues $\lambda(t)$, the energy levels $E_3$ and $E_3'$ can be extracted with the help of an 
exponential fit% to
\begin{gather}
    \lambda(t,t_0)\ \propto\ \left(e^{ - E_3 (t-t_0)}  + e^{ - E_3 (T - (t-t_0))}\right)    \,,\\
    \lambda'(t,t_0)\ \propto\ \left(e^{ - E_3' (t-t_0)}  + e^{ - E_3' (T - (t-t_0))}\right)    \, .
\end{gather}
We check that our results are stable by comparing 
the extracted energy levels to the case where we also include
% to results obtained by including 
the operators 
$\phi_1\phi_0^\dagger\phi_0$ and $\phi_0^3\phi_0^\dagger\phi_0$ in the correlator matrix.

All our ensembles have time extent $T=64$, and we have checked that this allows us to neglect all the uncertainties due to finite $T$ safely.
To reduce the statistical error of the correlators, we use translational invariance and average over all possible combinations with the same source-sink separation e.g.
\begin{equation}
\langle \tilde\phi_{0}^\dagger(t)\tilde\phi_{0}(0) \rangle=\frac{1}{T}\sum_{t'=0}^T\langle \tilde\phi_{0}^\dagger(t')\tilde\phi_{0}(t-t') \rangle\,. 
\end{equation}

For the implementation of the GEVP we use \cite{hadron:2020}, and
the values of the energies measured in our ensembles are reported in Table~\ref{tab:latt-spectrum}.

\begin{table}[t]
\footnotesize
\addtolength{\tabcolsep}{-0.5mm}
\renewcommand{\arraystretch}{1}
\centering
\begin{tabular}{ccllll}
\hline
\(g\) & \(L\) & \(M_0\) & \(E_2\) & \(E_3\) & \(M_1\) \\
\hline
0 & 20 & 0.246589(45) & 0.497378(84) & 0.75174(23) & 0.747951(77) \\
0 & 21 & 0.246538(47) & 0.496854(89) & 0.75021(23) & 0.748082(64) \\
0 & 22 & 0.246463(30) & 0.496035(81) & 0.74818(23) & 0.747621(70) \\
0 & 23 & 0.246425(61) & 0.49556(11) & 0.74718(21) & 0.747980(72) \\
0 & 24 & 0.246427(37) & 0.495346(78) & 0.74666(21) & 0.747911(74) \\
\hline
\(g\) & \(L\) & \(M_0\) & \(E_2\) & \(E_3\) & \(E_3'\) \\
\hline
4.43 & 20 & 0.272882(39) & 0.549142(71) & 0.82867(27) & 0.824666(91) \\
4.43 & 21 & 0.272865(37) & 0.548792(81) & 0.82731(27) & 0.824798(82) \\
4.43 & 22 & 0.272781(36) & 0.548287(62) & 0.82601(20) & 0.824666(83) \\
4.43 & 23 & 0.272684(45) & 0.547683(97) & 0.825158(97) & 0.82440(14) \\
4.43 & 24 & 0.272728(27) & 0.54766(10) & 0.82494(12) & 0.82398(23) \\
4.43 & 25 & 0.272718(39) & 0.54734(10) & 0.82478(10) & 0.82360(16) \\
4.43 & 26 & 0.272695(42) & 0.546890(85) & 0.824727(71) & 0.82268(24) \\
\hline
8.87 & 21 & 0.273556(45) & 0.550336(92) & 0.82968(24) & 0.826412(97) \\
8.87 & 22 & 0.273585(55) & 0.549907(67) & 0.82895(17) & 0.82650(14) \\
8.87 & 23 & 0.273541(46) & 0.54927(11) & 0.82748(16) & 0.82598(15) \\
8.87 & 24 & 0.273491(38) & 0.548826(77) & 0.82682(15) & 0.82580(28) \\
8.87 & 25 & 0.273515(41) & 0.548847(91) & 0.82669(13) & 0.82550(18) \\
8.87 & 26 & 0.273469(45) & 0.548479(93) & 0.826640(77) & 0.82493(12) \\
\hline
17.81 & 21 & 0.271285(32) & 0.54545(11) & 0.82281(19) & 0.818823(99) \\
17.81 & 22 & 0.271284(43) & 0.545207(85) & 0.82178(19) & 0.81890(11) \\
17.81 & 23 & 0.271233(50) & 0.544728(95) & 0.82070(19) & 0.81888(11) \\
17.81 & 24 & 0.271245(22) & 0.544502(85) & 0.82028(15) & 0.81843(16) \\
17.81 & 25 & 0.271249(38) & 0.54422(10) & 0.819879(93) & 0.81845(15) \\
17.81 & 26 & 0.271295(39) & 0.544079(73) & 0.819881(72) & 0.81807(28) \\
17.81 & 27 & 0.271131(38) & 0.543634(71) & 0.81976(15) & 0.81733(23) \\
\hline
\end{tabular}
\caption{
Energy levels determined in this work. $M_0$ is the one-particle mass and $E_2$ is the two-particle energy. When $g=0$, $M_1$ labels the mass of the heavy particle, and $E_3 $ the lowest three-particle energy level. When $g>0$, $E_3 $  and $E'_3$ correspond to two different three-particle energy levels. All energies are given in units of the lattice spacing.  
}
\label{tab:latt-spectrum}
\end{table}

%%%%%%%%%%%%%%%%%%%%%%%%%%%%%%
\section{Quantization Conditions}
%%%%%%%%%%%%%%%%%%%%%%%%%%%%%%
\label{sec:QCs}

We now describe the two- and three-particle quantization conditions. In the two-particle case, this is the well-established Lüscher formalism~\cite{LUSCHER1991531}. In the $S$-wave only limit, the two-particle spectrum is given by solutions of the following equation
\begin{gather}
             q^* \cot \delta = \frac{2}{ \gamma L  \sqrt{\pi} }  \mathcal Z_{00}^{\bm P}\left(1,\frac{q^* L}{2\pi} \right) 
             \,.\label{eq:luescher_qc}
\end{gather}
 Here  $\mathcal Z_{00}^{\bm P}$ is the Lüscher zeta function in the moving frame, the Lorentz boost factor ${\gamma=E_2( \bm{P})/E_{CM}}$  is defined in terms of center of mass energy $E^2_\text{CM}=E^2_2(\bm{P})-{\bm{P}}^2$ with $\bm P$ the total momentum of the system, while the relative momentum of the two particles is given by ${q^*}^2 = E_{CM}^2/4-M_0^2$.
 The right-hand side of Eq.~(\ref{eq:luescher_qc}) can be computed from the spectrum at a finite volume while the left-hand side is related to the scattering amplitude in the infinite volume.
 All the data quoted in this paper are in the $\bm P=0$ frame, $\gamma=1$, and we only consider the trivial irreducible representation $A_1$ of the octahedral group. 

As mentioned above, we utilize both the RFT and the FVU approach in the three-particle sector.
The RFT and FVU approaches have been shown to be formally equivalent, and a non-trivial integral-equation-type relation between the three-particle interaction parameters has been established~\cite{Blanton:2020jnm, Brett:2021wyd,Jackura:2022gib}. All approaches use scheme-dependent quantities to parametrize three-body effects: $\mathcal K_\text{df,3}$ for RFT and $C$ for FVU. The schemes differ, for example, due to different implementations of the spectator-momentum cutoff or different forms of the one-particle exchange terms. In the following, we recap both approaches in the case of three identical scalars and no two-to-three processes.

%%%%%%%%%%%%%%%%%%%%%%%%%%%%%%
\subsection{Relativistic Field Theory approach}
\label{sec:RFT}
%%%%%%%%%%%%%%%%%%%%%%%%%%%%%%

The RFT approach~\cite{Hansen:2014eka,Hansen:2015zga} is derived by classifying all power-law finite-volume effects emerging from all Feynman diagrams to all orders in perturbation theory in a generic relativistic field theory. 
In the case of three identical scalars with mass $M_0$ and no two-to-three transitions, the quantization condition reads
\begin{equation}
    \det \left[F_{3}(E_3,\bm{P},L) + 1/{ \mathcal K}_\text{df,3}(E_3^*)\right]=0\,,
    \label{eq:RFT}
\end{equation}
where $E^*_3 = \sqrt{E_3^2 - \bm P^2}$ is the center-of-mass (c.m.) three-particle energy, and $F_3$ and ${\mathcal K}_\text{df,3}$ are matrices in a space labeled by the finite-volume momentum, $\bm p=\frac{2\pi}{L}\bm n$, $\bm n \in\mathds{Z}^3$, of one of the particles (denoted the “spectator”) and the angular momentum of the other two in their two-particle c.m. frame. The above determinant acts in this space.
In the simplest case, with only $S$-wave interactions, taking the zero total momentum $\bm{P}=\bm{0}$ and assuming ${ \mathcal K}_\text{df,3}$ to be only a function of the overall energy of the system $E_3$, the quantization condition can be reduced to
\begin{equation}
F_{3}^\text{iso}(E_3,L)=-1/{ \mathcal K}_\text{df,3}^\text{iso}(E_3)\,. \label{eq:qc3iso}
\end{equation}
This is usually referred to as the isotropic approximation, which neglects higher partial waves in the three-particle system and two-particle subsystems, and is expected to be valid close to
the threshold. The right-hand side of this equation is the three-particle K-matrix, and it parametrizes three-particle short-range interactions. 
Note that ${ \mathcal K}_\text{df,3}^\text{iso}$ is a scheme-dependent unphysical  object. The connection to the physical amplitude will be discussed below in section~\ref{sec:infinitevolume}.

The left-hand side of Eq.~(\ref{eq:qc3iso}) contains finite-volume information and the two-particle scattering phase shift. The relevant expressions to compute $F_3^\text{iso}$ are:
%%%%%%%%
\begin{align}
\begin{split}
F_{3}^\text{iso}(E_3, L) &=\frac{1}{L^3} \sum_{\bm{kp}}\left[ \frac{\tilde F^s}{3}-
\tilde F^s\frac{1}{( \tilde{\mathcal{K}}_2^s)^{-1}+\tilde F^s+\tilde G^s}\tilde F^s\right]_{\bm{kp}}\,,
\\
\left[\tilde{\mathcal K}_2^s\right]_{\bm{kp}} &= \delta_{\bm{kp}}  \frac{32\pi\omega_{k} E_{2,{k}}^*}{ \left(q_{2,k}^{* } \cot\delta+|q^*_{2,k}|\left(1-H({k})\right)\right) }\,, %\left[(q_{2,k}^{*\, 2} \cot\delta )+|q^*_{2,k}|(1-H({k}))\right]
\\
\left[\widetilde F^s\right]_{\bm{kp}} &= \delta_{\bm{k p}} 
 \left(\frac1{L^3}\sum_{\bm a}^{\text{UV}} - \PV\int_{\bm a}^{\text{UV}} \right) \frac{ H({k})  }
{ 4\omega_{ k}4\omega_{ a} \omega_{ k+a}(E_3-\omega_{ k}-\omega_{ a}-\omega_{k+a})}\,, 
%\label{eq:Fsa}, 
\\
\left[\widetilde G^s\right]_{\bm{kp} } &= \frac{H({k}) H({p})}{L^3 2 \omega_{ k} 2 \omega_{ p} \left( ( P-p-k )^2-M_0^2 \right)}
\,.
\label{eq:eqsRFT}
\end{split}
\end{align} 
%%%%%%%%
Here, the vectors $\bm{p}$ and $\bm{k}$ label the finite-volume momenta of the spectator particle, and $\tilde{\mathcal K}_2^s, \widetilde F^s$ and $\widetilde G^s$ are matrices in a space with $\bm{p}, \bm{k}$ indices. The on-shell energies for particles with momentum $\bm x$ are denoted by $\omega_x=\sqrt{\bm{x}^2+M_0^2}$, while the c.m. energy of the interacting pair and relative c.m. momentum are given by ${E_{2,k}^{*\, 2} = (E_3-\omega_{ k})^2- \bm{k}^2 = E_3^2+M_0^2 - 2 E_3 \omega_{ k}}$, $q_{2,k}^{*\, 2} = {E_{2,k}^{*\,2}}/{4} - M_0^2$. Moreover, $k$ and $p$ are the four momenta of the spectator particles and $P=(E_3, 0)$ in the overall c.m. frame.
In $\widetilde F^s$, the integral is defined $\int_{\bm a} \equiv d^3 a/(2\pi)^3$, while the sum over $\bm a$ runs over all finite-volume momenta.
The principal value (PV) prescription is defined as in  Ref.~\cite{Hansen:2014eka}.
The superscript ``UV'' in the sum and integral indicate that an ultraviolet cutoff is required to separately evaluate the sum and integral. A method for evaluating numerically the sum minus integral can be found in appendix B of Ref.~\cite{Briceno:2018mlh}.

The expressions in Eqs.~(\ref{eq:eqsRFT}) contain a smooth cutoff function $H({k})$  as defined in Eqs.~(28) and (29) of Ref.~\cite{Hansen:2014eka}, which we display here for completeness
%%%%%%%%
\begin{align}
\label{eq:Hdef}
H(k) & = J\left(\frac{E_{2,k}^{*2} - (1+\alpha) M_0^2}{(3-\alpha)M_0^2}\right)\,,\quad\, %\\
%z&= \frac{E_{2,k}^{*2} - (1+\alpha) m^2}{(3-\alpha)m^2} \,,
J(z)= 
\begin{cases}
0 \,, & z \le 0 \,; \\ \exp \left( - \frac{1}{z} \exp \left
[-\frac{1}{1-z} \right] \right ) \,, & 0<z < 1 \,; 
\\ 1 \,, & 1\le z
\,.
\end{cases}
\end{align}
%%%%%%%%
In this work, we keep $\alpha=-1$, which ensures that all matrices appearing in the quantization condition are finite. In particular, the cutoff function restricts $\bm k < \bm k_\text{max}$, where 
$\bm k_\text{max}$ is defined by $E_{2,k}^{*\, 2}\big \rvert_{ k_\text{max}}  = 0$. We use the implementation of the quantization condition provided in Ref.~\cite{Blanton:2021eyf} and the associated repository~\cite{QC3git}.

%%%%%%%%%%%%%%%%%%%%%%%%%%%%%%
\subsection{Finite-Volume Unitarity approach}
\label{sec:FVU}
%%%%%%%%%%%%%%%%%%%%%%%%%%%%%%

The FVU approach is based on the unitarity relations for the three-to-three body scattering amplitude. Hereby, the bookkeeping of various configurations of three particles, going on-shell, is simplified by utilizing the so-called isobar-spectator language~\cite{Mai:2017vot}. The isobar can be thought of as an intermediate auxiliary field which, in particular, can also describe a system of two repulsively interacting 
particles~\cite{Mai:2018djl, Culver:2019vvu, Alexandru:2020xqf}.
The notion of an isobar is closely related to the
particle pair in the RFT approach -- namely, a full propagator of an isobar coincides with the two-particle Green function in a particular partial-wave channel.
Using this amplitude and employing constraints on intermediate momenta due to the (periodic) boundary conditions on the lattice, this approach yields the  
FVU three-body quantization condition. In the c.m. frame, it is algebraically identical to the Lorentz-invariant NREFT quantization condition as mentioned before, and hence, if needed, one could use the same procedure to transform the equation to the moving frames. In this paper, however, we work explicitly in the c.m. frame and the need for such a transformation does not arise. Alternatively, one might consider the ``relativization'' of the one-particle
exchange term similar to the last line in Eq.~\eqref{eq:eqsRFT} and setting a cutoff low enough to ensure that no spurious energy levels emerge in the spectrum.

For the present case with only $S$-wave interactions the FVU three-body quantization condition reduces to
%%%%%%%%%%%%%
\begin{align}
 &{\rm det}\left[B+C+E_L\cdot\left(\tilde K^{-1}-\Sigma^{FV}\right)\right]_{\bm{k}\bm{p}} = 0 \, .
\label{eq:FVU-3bQC}
\end{align}
%%%%%%%%%%%%%
Explicitly, the above matrices in the space of spectator momenta are defined as
%%%%%%%%%%%%%
\begin{align}
\label{eq:FVU:B}
\left[B(E_3)\right]_{\bm{k}\bm{p}}&=B(\bm{k},\bm{p};E_3)=\frac{-1}{2\omega_{{k}+{p}}(E_3-\omega_{{k}}-\omega_{{p}}-\omega_{{k}+{p}})}\,,\nonumber\\
%&\left[C(E_3)\right]_{\bm{p}\bm{p}'}=\frac{c_ 0}{E_3^2-m_1^2}+c_1+c_2E_3^2\,,\nonumber\\
\left[{\tilde K^{-1}}(E_3)\right]_{\bm{k}\bm{p}}&=\delta^3_{\bm{k}\bm{p}}{\tilde K}^{-1}_{2}\left(E_{2,{p}}^{*\, 2}\right)\,,\\
\left[\Sigma^{FV}(E_3)\right]_{\bm{k}\bm{p}}&=\delta^3_{\bm{k}\bm{p}}\Sigma^{FV}(E_{2,{p}}^{*\, 2},M_0L,\bm{p})\,,\nonumber\\
\left[E_L\right]_{\bm{k}\bm{p}}&=\delta^3_{\bm{k}\bm{p}}(M_0L)^32\omega_{{p}}\,.\nonumber
\end{align}
%%%%%%%%%%%%%
Here, we used the same nomenclature as introduced in the previous section. Ignoring the exponentially suppressed $e^{-M_0L}$ terms, the only volume-dependent terms are given by the kinematical function $E_L$, the one-particle exchange diagram with propagator $B$, and the two-body self-energy term $\Sigma^{FV}$ evaluated in the finite volume. The form of the latter two is fixed by ensuring two- and three-body unitarity in the infinite volume as discussed before. In particular, the two-body self-energy term reads 
%%%%%%%%%%%%%
\begin{align}
    % &\Sigma^{FV}(\sigma,M_0L,\bm{p})=\frac{J_{\bm p}(\sigma)}{(M_0L)^3}
    % \sum_{\bm{s}}
    % % \sum_{\bm{k}\in \mathcal{S}_L}
    % \frac{\sigma}{(4\omega_{{s}^\star}^2)}\frac{1}{2\omega_{{s}^\star}}\frac{1}{\sigma-4\omega_{{s}^\star}^2}\,
    &\Sigma^{FV}(E^{*2}_{2,p},M_0L,\bm{p})=\frac{J_{\bm p}(E^{*2}_{2,p})}{(M_0L)^3}
    \sum_{\bm{s}}
    % \sum_{\bm{k}\in \mathcal{S}_L}
    \frac{E^{*2}_{2,p}}{(4\omega_{{s}^\star}^2)}\frac{1}{2\omega_{{s}^\star}}\frac{1}{E^{*2}_{2,p}-4\omega_{{s}^\star}^2}\,
\end{align}
%%%%%%%%%%%%%%%%%%%%%%%%%%
and Lorentz boost with the three-momentum $\bm{p}$,
%%%%%%%%%%%%%%%%%%%%%%%%%%
\begin{align}
    \bm{s}^\star=
    \bm{s}+\bm{p}\left(\frac{\bm{s}\cdot\bm{p}}{\bm{p}^2}\left(J_{\bm p}(E^{*2}_{2,p})-1\right)+
    \frac{J_{\bm p}(E^{*2}_{2,p})}{2}\right)\,,~~
    J_{\bm p}(E^{*2}_{2,p})=\sqrt{\frac{E^{*2}_{2,p}}{E^{*2}_{2,p}+\bm{p}^2}}\,
    ,
\end{align}
%%%%%%%%%%%%%
see Ref.~\cite{Mai:2019fba}. The matrix $C$ in the isotropic approximation is a matrix in the spectator momenta, where all entries are identical and depend only on the total energy $E_3$. It is a
volume-independent term that, together with $\tilde K_2^{-1}$, encodes the three- and two-body dynamics, respectively.
Hence, they cannot be fixed from principles of the S-matrix theory alone, but only from the fits to the actual finite-volume spectra. We note that
fixing $C$ requires the calculation of the three-body spectrum, whereas $\tilde K_2^{-1}$ can be fixed either from the two-body spectrum alone or from a combined fit of the two- and three-body energy levels.
Specifically, using the standard L\"uscher approach~\eqref{eq:luescher_qc} we express the two-body term as
%%%%%%%%%%%%%
\begin{align}
\label{eq:FVU:Kmatrix}
    % k\cot \delta = -16\pi\sqrt{\sigma}\left(\tilde K^{-1}(\sigma)-\Re \Sigma^{IV}(\sigma)\right)\,,
    % \tilde K^{-1}_2(\sigma)=\frac{-q^*\cot \delta}{16\pi\sqrt{\sigma}}+\Re \Sigma^{IV}(\sigma)\,,
    \tilde K^{-1}_2(E^{*2}_{2,p})=\frac{-q^{*,2}_{2,p}\cot \delta}{16\pi\sqrt{E^{*2}_{2,p}}}+\Re \Sigma^{IV}(E^{*2}_{2,p})\,,
\end{align}
%%%%%%%%%%%%% 
through the two-body $S$-wave phase-shift $\delta$ with $q^{*,2}_{2,p}$ defined as in the previous section. The relative c.m. momentum is defined as in section~\ref{sec:RFT},
%is equal to $q^*=\sqrt{\sigma/4-M_0^2}$,
and the infinite-volume self-energy reads
%%%%%%%%%%%%%
\begin{align}
\label{eq:FVU:SigmaIV}
    % \Sigma^{IV}(\sigma)=\int\frac{d^3\bm{s}}{(2\pi)^3}
    % \frac{1}{2\omega_{{s}}}
    % \frac{\sigma}{(4\omega_{{s}}^2)}\frac{1}{\sigma-4\omega_{{s}}^2}\,.
    \Sigma^{IV}(E^{*2}_{2,p})=\int\frac{d^3{s}}{(2\pi)^3}
    \frac{1}{2\omega_{{s}}}
    \frac{E^{*2}_{2,p}}{(4\omega_{{s}}^2)}\frac{1}{E^{*2}_{2,p}-4\omega_{{s}}^2}\,.
\end{align}
%%%%%%%%%%%%%%%%%%%%%%%%%%
%This integral can indeed be evaluated using complex-contour deformation discussed in section~\ref{sec:IVU}. 
We note that other parametrizations of the two-body term $\tilde K^{-1}_2$ can be chosen as well.

Finally, we note that all available three-body quantization conditions, Eqs.~(\ref{eq:FVU-3bQC}) and (\ref{eq:RFT}), are infinite-dimensional in the spectator momentum space, which calls for a truncation of the momentum space. Various approaches to this issue have been discussed in the literature, such as the inclusion of the form-factors~\cite{Hansen:2015zga, Mai:2017bge, Mai:2019fba}, hard cutoff, or over-subtractions~\cite{Alexandru:2020xqf, Mai:2021nul, Sadasivan:2020syi}. These schemes all come with various (dis)advantages. Here we work with a hard cutoff $|\bm{p}|<\Lambda$ with $\Lambda=\sqrt{8}\pi/L$, which is sufficient to access the kinematical region of interest, see section~\ref{sec:results}.

%%%%%%%%%%%%%%%%%%%%%%%%%%%%%%%%%%%%%%%%%%%%%%%%%%
\subsection{Parametrization of the two- and three-body forces}
\label{subsec:fitting}
%%%%%%%%%%%%%%%%%%%%%%%%%%%%%%%%%%%%%%%%%%%%%%%%%%

Finding the solutions of the quantization conditions in Eqs.~\eqref{eq:luescher_qc},~\eqref{eq:qc3iso}, and~\eqref{eq:FVU-3bQC} 
%Eqs.~\eqref{eq:luescher_qc} in the two-particle sector, and either Eq.~\eqref{eq:qc3iso} or Eq.~\eqref{eq:FVU-3bQC} in three-particle sector,
allows one to predict the energy eigenvalues, given the knowledge of volume-independent quantities. This also means that we can extract these volume-independent quantities from finite-volume spectra. In particular, we are interested in constraining $q^*\cot\delta$ for the two-body sector, and $\Kiso$ (RFT) and $C$ (FVU) in the three-body sector.

In practice, we need to parametrize the energy dependence of the quantities, describing the interactions, with a small set of parameters. For the two-particle interactions, it is customary to use the effective range expansion:
%%%%%%%
\begin{equation}
   q^* \cot\delta = \frac{1}{ a} + \mathcal{O}({q^*}^2)\,,
\end{equation}
%%%%%%%
where $a$ is the scattering length. In the three-body sector, we will use parametrizations that include an explicit pole to accommodate a resonance. In particular, $C$ in FVU and $\Kiso$ in RFT will be parametrized as
%%%%%%%
\begin{align}
    C=\frac{c_0}{E_3^{2}-m_R^2} +c_1\,,\qquad
   \Kiso=\frac{c_0'}{E_3^{2}-m_R'^2} +c_1' \,,
   \label{eq:3bf-FVU/RFT}
\end{align}
%%%%%%%
where $c_i$ and $c'_i$ ($i=0,1$) are numerical constants to be determined from the data. When the final-state rescattering is weak, one can relate the sign of $c_0$ ($c_0'$) to the residua of the two-point correlation function. The latter has a definite sign as discussed in Ref.~\cite{Blanton:2019igq}, which implies that $c_0 (c_0')$ should be negative.
%To ensure a physical residue, $c_0$, $c_0'$ have to be negative, at least in the assumption that weak final-state rescattering is unable to change the sign of the residue.

Previous studies \cite{Mai:2021nul, Briceno:2018mlh} indicate that these parametrizations can describe resonances. 
Moreover, it should be pointed out that the parametrization given in
Eq.~(\ref{eq:3bf-FVU/RFT}) is already rather general. For example, on physical grounds, one
may exclude a double pole or a cut in the variable $E_3^2$. The latter can be directly ruled out from unitarity considerations in the low-energy region. Regarding the former, it is expected that the weak repulsive final-state interactions will lead to a small splitting of the double pole, resulting in two nearby poles in the scattering amplitude -- an implausible scenario in the model studied in this work. We, therefore, refrain from considering these rather exotic scenarios and concentrate on the simple parametrization of Eq.~(\ref{eq:3bf-FVU/RFT}). The only freedom left in this expression is adding polynomial terms in $E_3^2$ to the background or adding more poles. It is also worth mentioning that, under certain circumstances, even a three-body force without poles can lead to a dynamical generation of resonances (see Ref.~\cite{Mai:2021nul}). In this case, the background would be described by a higher-order polynomial which mimics the Taylor expansion of the pole term at low-energy. Unlike Ref.~\cite{Mai:2021nul}, our model produces weakly repulsive two-particle interactions, and so, a dynamical generation of poles is not expected. In addition, fits to a higher-order polynomials are very unstable. For these reasons, we opt to simply use the model in Eq.~(\ref{eq:3bf-FVU/RFT}), which as will be seen, will provide a good description of the data. 

Since the two-particle subsystem is not resonant, finding solutions of Eq.~\eqref{eq:luescher_qc} is straightforward. In contrast, the presence of a resonance in the three-particle spectrum could make the problem numerically unstable. To ameliorate this problem, we multiply the quantization conditions by the denominator of the three-body force. Thus, the quantization condition becomes
%%%%%%%
\begin{align}
    % &(E_3^3-m_1^2) (\Sigma^{FV}(E_2^2,M_0L,\bm{P}_2)-{\tilde K}^{-1}(E_2^2-\bm{P}_2^2))=0\,,\\
    (E_3^3-m_R^2)\, {\rm det}\left[B+C+E_L\cdot\left(\tilde K^{-1}-\Sigma^{FV}\right)\right] \!\!&= 0\,,\\[2mm]
   (E_3^3-m_R'^2)\,(1/F_{3}^\text{iso}(E_3,L)+{ K}_\text{df,3}^\text{iso}(E_3))&=0 \,.
\end{align}
%%%%%%%
Given the parametrizations in Eqs.~(\ref{eq:3bf-FVU/RFT}), the solution of the above quantization condition will give the predicted energy levels. These modified quantization conditions do not have a pole  at $E=m_R$ for FVU or $m_R'$ for RFT. However, in the case of $c_0$ or $c_0'$  equal to zero, they will both have roots, describing a three-particle system with constant three-body force, and one stable particle with no finite-volume effects and constant mass $m_R$ or $m_R'$. 

The final step involves extracting the parameters by performing a fit to the energy levels.  This is the so-called ``spectrum method'' see, e.g. Refs.~\cite{Morningstar:2017spu, Blanton:2021llb} . In our case, we simultaneously fit the two- and three-particle spectra, finding the values of the parameters $p_n$ such that the correlated $\chi^2$-function becomes minimal:
%%%%%%%
\begin{equation}
    \chi^2= \sum_{i,j}   \left(  E_{i}(p_n)  - E_{i}^\text{data}\right) \tilde C^{-1}_{ij} \left(  E_{j}(p_n)  - E_{j}^\text{data}\right)\,.
\end{equation}
%%%%%%%
Here $\tilde C$ is the covariance matrix of the lattice energy levels $E_{i}^\text{data}$ in the two- and three-particle spectrum. Moreover, $E_{i}(p_n)$ are the predicted energy levels obtained by solving the quantization conditions with the given  parameters.

%%%%%%%%%%%%%%%%%%%%%%%%%%%%%%%%%%%%%%%%%%%%%%%%
\section{Analysis of finite-volume spectra}
\label{sec:results}
%%%%%%%%%%%%%%%%%%%%%%%%%%%%%%%%%%%%%%%%%%%%%%%%
In this section, we present our numerical results. To summarize, we observe that FVU and RFT lead to qualitatively identical data descriptions, i.e., the best fit with both formalisms gives similar $\chi^2$ (see Table~\ref{TAB:2&3body-fits}).
%and the physical quantities, as the mass or with of the resonance, extracted with both formalism agrees within error (Sec.~\ref{subsec:pole_position}). 
After the numerical demonstration of the equivalence of the two formalisms, we present an investigation of our model in the limit of zero coupling $g$ and a check that the scenario without a pole in the three-particle amplitude is not compatible with our data (section~\ref{subsec:no_resonance_fit}).

%%%%%%%%%%%%%%%%%%%%%%%%
%%%%%%%%%%%%%%%%%%%%%%%%
\begin{table}[t]
\footnotesize
% \hspace{0cm}
\addtolength{\tabcolsep}{-0.5mm}
\renewcommand{\arraystretch}{1}
%\begin{tabular}{|c| c | ccccccc|c|}
\begin{tabular}{c c  cccccccc}
    %%%%%%%%%%%%%%%%%%%%%%%%%%%%%%
    %%%%%%%%%%%%%%%%%%%%%%%%%%%%%%
    \hline
\,$g$\,&   
    &$a\,M_0$ & $m_R/M_0$& $c_0$ &$c_1M_0^2$
    & $m_R'/M_0$& $c_0'$ &$c_1'M_0^2$ &$\chi^2_{\rm dof}$\\
    \hline
\multirow{4}{*}{4.43}& FVU
    &$-0.1512(09) $& $3.0229(1) $& $-0.0188(35)$& $-$ & $-$& $-$& $-$ &$2.9$\\%[9/7/22]
    &RFT &$-0.1522(12)$&$-$ & $-$& $-$& $3.0232(2)$& $-31.6(8.4)  $&$-$& $2.5$\\  \cline{2-10}
     & FVU
     &$-0.1569(12)  $& $3.0233(2) $&$-0.0297(57)$&$2.29(38)$&$-$ & $-$& $-$&$1.5$\\
     &RFT&$-0.1571(10)$ &$-$ & $-$& $-$& $3.0237(2)$&$-37.6(9.0) $&$  2789(540) $& $1.5$\\
     \hline 
 \multirow{4}{*}{8.87}& FVU
     &$-0.1521(11) $& $3.0205(2) $& $-0.0475(66)$&$-$&$-$ & $-$& $-$&$1.7$\\
     &RFT&$-0.1531(13)$&$-$ & $-$& $-$& $3.0212(3) $& $-80(14)$ &$ -$ &1.6\\\cline{2-10}
    &FVU&
    $-0.1549(16) $& $ 3.0205(2) $& $-0.0595(99)$&$0.93(41)$&$-$ & $-$& $-$&$1.5$\\
    &RFT&$-0.1563(27)$&$-$ & $-$& $-$& $ 3.0213(3)$& $-97(16)$ & $1773(980)$ &$1.4$\\
    \hline
 \multirow{4}{*}{17.81}& FVU
    &$-0.1444(11)  $& $3.0184(2) $& $-0.1136(77)$&$-$&$-$ & $-$& $-$&$1.6$\\
    &RFT&$-0.1450(17) $&$-$ & $-$& $-$& $3.0199(2)$& $-178(17)$ & $-$ &$1.6$\\\cline{2-10}
    &FVU 
    &$-0.1464(14) $& $3.0183(2) $& $-0.1363(148)$&$0.84(39)$&$-$ & $-$& $-$&$1.3$\\
    &RFT&$-0.1484(16)$  &$-$ & $-$& $-$&  $3.0200(2)$& $-210(23)$ & $2227(600)$ &$1.2$\\
    \hline
\end{tabular} 
\caption{Summary of the FVU and RFT fits to the two- ($E_2$) and three-body ($E_3$) levels, including (cross)correlations. For each bare coupling $g$, the results represent three and four-parameter fits, respectively.
}
\label{TAB:2&3body-fits}
\end{table}
%%%%%%%%%%%%%%%%%%%%%%%%
%%%%%%%%%%%%%%%%%%%%%%%%

%%%%%%%%%%%%%%%%%%%%%%%%
%%%%%%%%%%%%%%%%%%%%%%%%
\renewcommand{\textfraction}{0.01}
\renewcommand{\topfraction}{0.01}
\renewcommand{\bottomfraction}{0.01}
\renewcommand{\floatpagefraction}{0.01}
\setcounter{totalnumber}{1}
\begin{figure}[p!]
\begin{center}
    \includegraphics[width=0.7\textwidth, trim={0 1.31cm 0 0.0cm },clip]{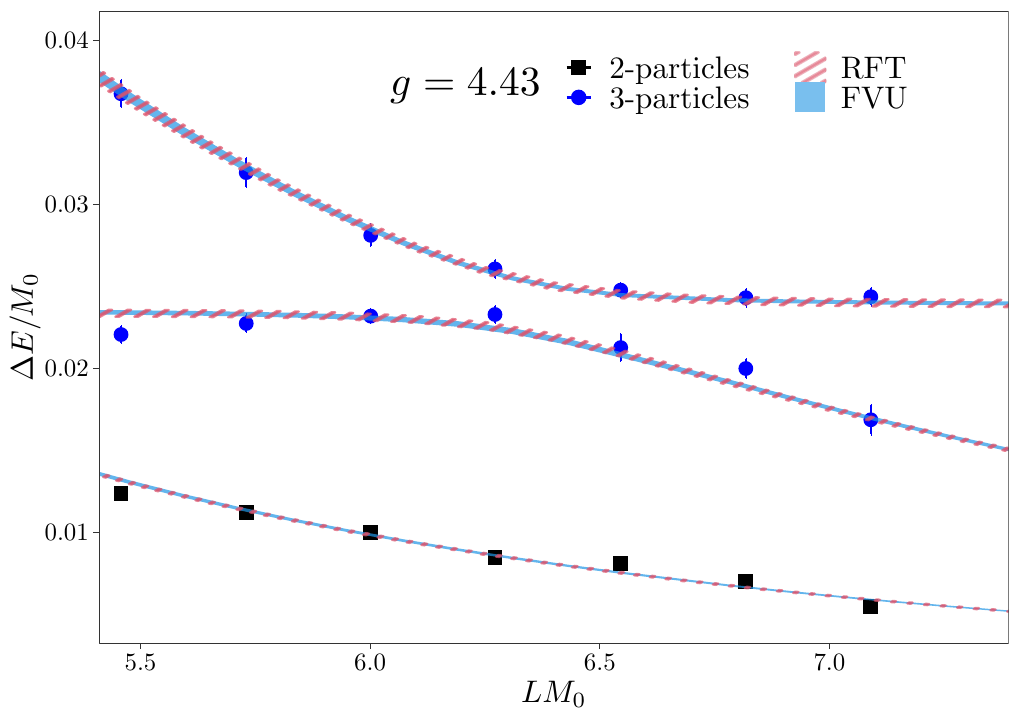}\\
    \includegraphics[width=0.7\textwidth, trim={0 1.31cm 0 0.25cm },clip]{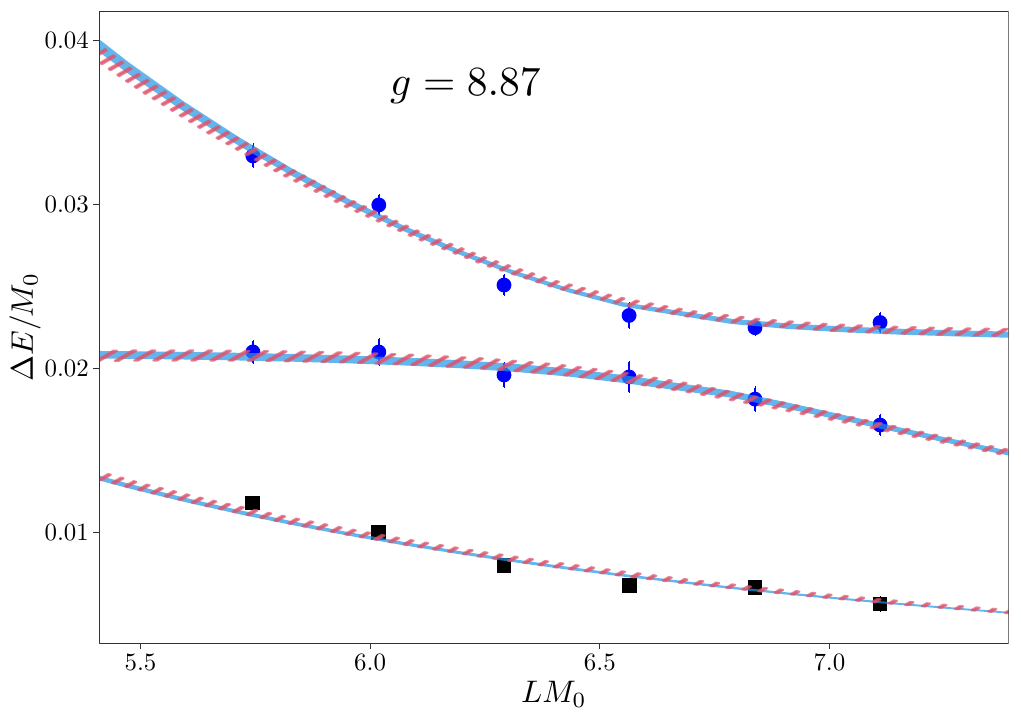} \\
    \includegraphics[width=0.7\textwidth, trim={0 0 0 0.25cm },clip]{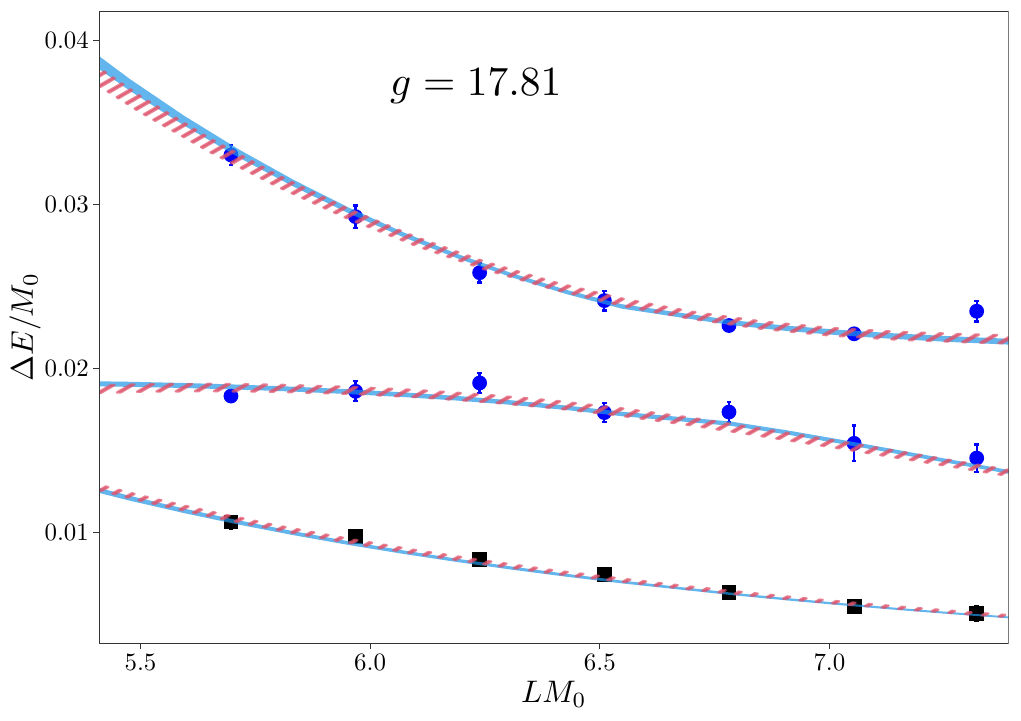}
\end{center}%\vspace{-0.5cm}
    \caption{The energy shift $\Delta E/M_0=E_j/M_0-j,~j=2,3$ of the two- and three-particle systems as a function of $LM_0$ for three values of $g$. 
    These data points are compared with the best fit results (with 4 parameters) of RFT (red stripes) and FVU (shaded blue) approaches to the energy levels, respectively. 
    }
    \label{fig:FVU_vs_QFT}
\end{figure}
\renewcommand{\textfraction}{0.3}
\renewcommand{\topfraction}{0.7}
\renewcommand{\bottomfraction}{0.5}
\renewcommand{\floatpagefraction}{0.5}
%%%%%%%%%%%%%%%%%%%%%%%%

%%%%%%%%%%%%%%%%%%%%%%%
\subsection{Numerical comparison of FVU and RFT}
\label{subsec:comparison_FVU_RFT}
%%%%%%%%%%%%%%%%%%%%%%%

We fit our models to the data, measured for our ensembles with $g=4.43, 8.87$ and $17.81$ (Table~\ref{tab:latt-spectrum}), as described in section~\ref{subsec:fitting}. Our best-fit results are reported in Table~\ref{TAB:2&3body-fits} and the resulting spectrum prediction is plotted in Fig.~\ref{fig:FVU_vs_QFT}. In that figure, the three panels correspond to the three non-zero values of the coupling $g$. In all three panels, we plot $\Delta E/M_0$ as a function of $LM_0$. Here, $\Delta E$ represents the energy shift in the two- (black squares) and three-particle (blue circles) systems, respectively ($\Delta E = E_j - j M_0$ with $j=2,3$ for the two- respectively three-particle systems). The bands represent our best fits to the data with the RFT (red stripes) and the FVU (shaded blue) parametrizations.

In the two-body sector, we obtain compatible results within the RFT and FVU approach for the scattering length $a$. Note, however, that in the three-body sector, the parameters are not directly comparable due to the scheme dependence discussed above. Nevertheless, we find that the parametrization used in RFT and the one in FVU Eq.~(\ref{eq:3bf-FVU/RFT}) can fit the data with good $\chi^2$ and they both give consistent predictions of the energy levels. We also performed a fit with and without the parameters $c_1$ or $c_1'$, which correspond to the background term in the three-body force. The inclusion of this extra parameter in the fit gives a small reduction of the $\chi^2$ in the cases of $g=8.87$ and $g=17.81$, while it is essential to fit the data at $g=4.43$.
We observe that $c_0'$ and $c_0$ are non-zero within errors and their mean values increase with the bare parameter $g$. This can also be appreciated in the spectrum: the avoided level crossing, which is characteristic of a resonance, becomes wider with increasing values of $g$ (see Fig.~\ref{fig:FVU_vs_QFT}).
The values for $m_R$ and $m_R'$ reported in Table~\ref{tab:latt-spectrum} are close to each other even if they are different between errors. The similarity may be due to the pole in the amplitude being very close to the real axis (see section~\ref{subsec:pole_position}), thus $m_R$ is not so far from the physical parameter $M_R$.

%%%%%%%%%%%%%%%%%%%%%%%%%%%%%%%%%%%%%%%
\subsection{Testing the resonance hypothesis}
\label{subsec:no_resonance_fit}
%%%%%%%%%%%%%%%%%%%%%%%%%%%%%%%%%%%%%%%

% %%%%%%%%%%%%%%%%%%%%%%%%
% \subsection{The model at $g=0$}
% \label{subsec:g0}
% %%%%%%%%%%%%%%%%%%%%%%%%

\begin{figure}[t]%[hbtp]
    \centering
    \includegraphics[width=0.49\linewidth]{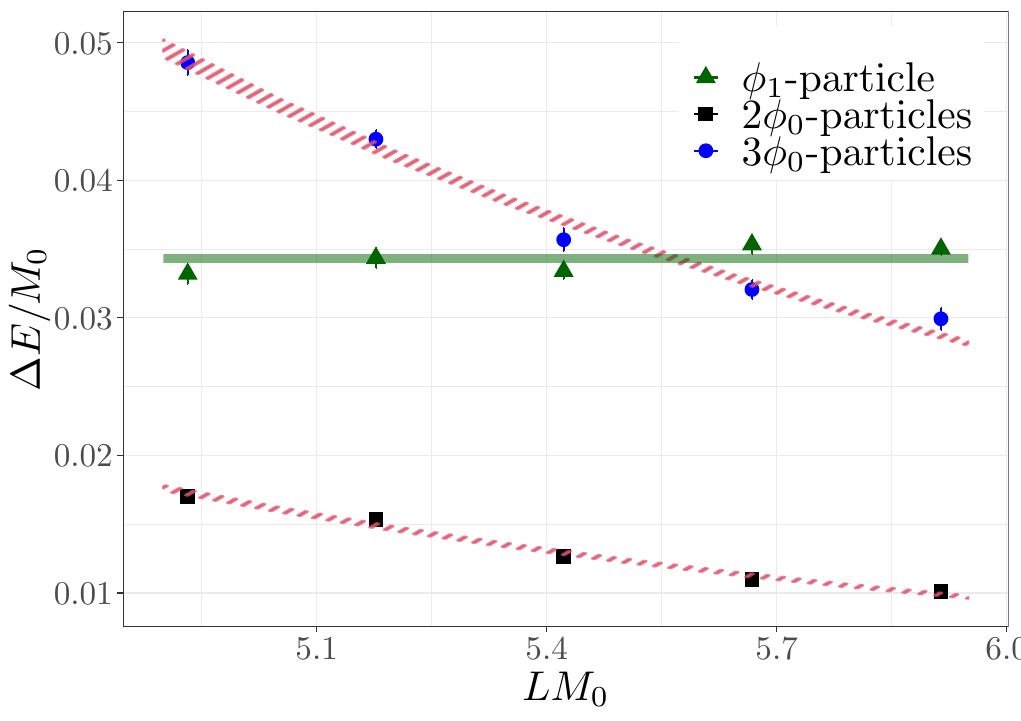}
    \includegraphics[width=0.49\linewidth]{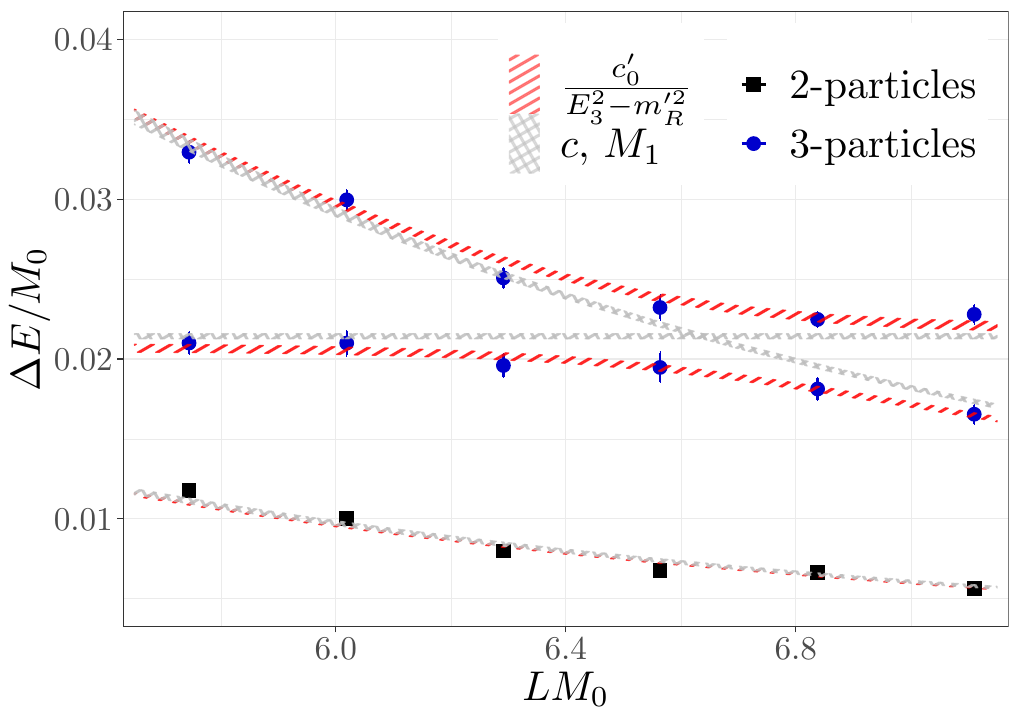}
     \caption{
    Interacting energy level shifts of two- (black squares) and three-particle (blue circles) systems, %with the free energies subtracted by $2M_0$ and $3M_0$ respectively and normalised by $M_0$ 
    as functions of $LM_0$.
    {\bf Left:} The case of $g=0$: Here, the $\phi_1$-particle is stable
%    and we subtracted the corresponding energy levels with $3M_0$  and normalised by $M_0$ as 
    (green triangles on the plot). 
    The green solid band represents the fit result of all the $\phi_1$ energy levels to a constant,
    %in $LM_0$, 
    while the red striped bands are the fit to the two- and three $\phi_0$-particle energy levels with the RFT quantization condition Eq.~(\ref{eq:const_fit}). 
    {\bf Right:} 
    In this panel we show results for $g=8.87$. We compare two fit models, where the first assumes the that $\phi_1$-particle is a resonance and the second assumes that it is stable. The red-striped bands represent the fit (\ref{eq:3bf-FVU/RFT}) with $c_1'$ set to zero,  reported in the sixth row of Table~\ref{TAB:2&3body-fits}, while the gray crosshatch band is the result of the fit  with $\Kiso$ constant and a constant energy level equal $M_1$. In both fits the two-particle sector is fitted with the Lüscher quantization condition with parametrization given in Eq.~(\ref{eq:luescher_qc}).
    }
    \label{fig:RFT_g0}
\end{figure}

In this section, we study the manifestation of a resonance in the finite-volume spectrum with its signature as an avoided level crossing. First, we consider the case of vanishing coupling $g$, i.e.,  when the  particle $\phi_1$ becomes stable and decoupled from $\phi_0$. 
This setting can be seen as a benchmark, since it corresponds to the well-studied $\varphi^4$ theory, and all scattering quantities are expected to be described by very simple parametrizations.
Our choice of $\kappa_0$ and $\kappa_1$ is such that the energy levels of the particle $\phi_1$ and three-particle $\phi_0$ cross around $LM_0\sim 5.6$ (Fig.~\ref{fig:RFT_g0}). Note that this crossing does not imply $\phi_1 \to 3\phi_0$ transitions, which are excluded based on the symmetries of the theory at $g=0$.
The energy level corresponding to one heavy particle $\phi_1$ can be simply extracted from the exponential fit at large time separation to the two-point correlation function Eq.~\ref{eq:M1}.
We fit the value of $M_1$ at each volume as a constant since we only expect exponentially suppressed finite-volume effects. The two and three $\phi_0$-particles are fitted with the RFT formalism with $\Kiso=c$ and 
$q^*\cot \delta= 1/a$. The best-fit values  are
\begin{align}\label{eq:const_fit}
    \chi^2_\text{dof}&=1.8  \\
    c\,M_0^2&=1351(490)\nn\\
    M_1/M_0&=3.03431(32)\nn\\
    a\,M_0&=-0.1514(18)\,.
\end{align}
This means that in the limit $g\to0$, the scenario of  $\phi_1$ as a stable particle decoupled from $\phi_0$ is supported by the data, and has a reasonably good fit quality, as expected.
%This means that in the limit $g\to0$, $\phi_1$ can be seen as a stable particle decoupled from $\phi_0$.
\footnote{It should be stressed once more that we did not include a pole term in the parametrization of the kernel, since the results at $g\neq0$ indicate that the residue of this term must vanish, as $g\to 0$. Furthermore, three particles with weak repulsive interactions are not expected to produce a shallow bound state and hence the energy level, which has almost no dependence on $L$, can be safely interpreted as a one-heavy-particle state. }

%%%%%%%%%%%%%%%%%%%%%%%%
%%%%%%%%%%%%%%%%%%%%%%%%
%%%%%%%%%%%%%%%%%%%%%%%%
%%%%%%%%%%%%%%%%%%%%%%%%

Now, we turn on the interaction between the $\phi_0$ and $\phi_1$ fields and repeat the above test. In particular, we want to check if the interpretation of $\phi_1$ as a stable particle would also be supported by data in this case at non-zero $g$-values. We do so by fitting 
either $\{c, M_1\}$ as before, or the form given in Eq.~\eqref{eq:3bf-FVU/RFT}.
The result of both fits can be found in  Fig.~\ref{fig:RFT_g0}, where the non-resonant fit is represented as the gray crosshatch band. As can be seen, the non-resonant fit fails to describe the data close to the avoided level crossing. The best-fit result is
\begin{align}
    \chi^2_\text{dof}&=3.1 \,,\nn \\
    cM_0^2&=-161\pm (880) \,,\nn \\ 
    m_R/M_0&=3.02142 (16) \,,\nn\\ 
    aM_0&=-0.1553(22) \nn\,. 
\end{align}
Note that the $\chi^2_\text{dof}$ of this fit is much worse than the one of the benchmark fit displayed in Eq.~(\ref{eq:const_fit}), which rules out this model for the scattering quatities.
 On the other hand, a fit with a pole in the $\Kiso$ matrix \eqref{eq:3bf-FVU/RFT} with two parameters (i.e. with $c_1=0$) gives a $\chi^2_{\rm dof}=1.6$ as reported in the sixth line of Table~\ref{TAB:2&3body-fits} and displayed in Fig.~\ref{fig:RFT_g0} as a solid red band.When including the pole, the  $\chi^2_\text{dof}$ is of the same order of the benchmark fit at $g=0$.

%%%%%%%%%%%%%%%%%%%%%%%%%%%%%%%%%%%%%%%
\section{Infinite-volume scattering}
\label{sec:infinitevolume}
%%%%%%%%%%%%%%%%%%%%%%%%%%%%%%%%%%%%%%%

After having determined the two- and three-body parameters, the goal is to extract physical resonance parameters, namely, the resonance pole position. However, a technical complication in the three-particle finite-volume formalism(s) is that the three-body parameters, $\Kiso$ or $C$, are scheme-dependent and therefore unphysical. In order to remove the scheme dependence, a set of integral equations leading to the physical scattering amplitude
needs to be solved. To do so, we use the state-of-the-art tools that have been developed separately for each method,  Refs.~\cite{Doring:2009yv, Sadasivan:2021emk, Hetherington:1965zza} for the FVU approach, and Refs.~\cite{Jackura:2020bsk,Hansen:2020otl} for the RFT.

%%%%%%%%%%%%%%%%%%%%%%%%
%%%%%%%%%%%%%%%%%%%%%%%%
\begin{figure*}[t]
    \centering
    \includegraphics[width=0.49\linewidth]{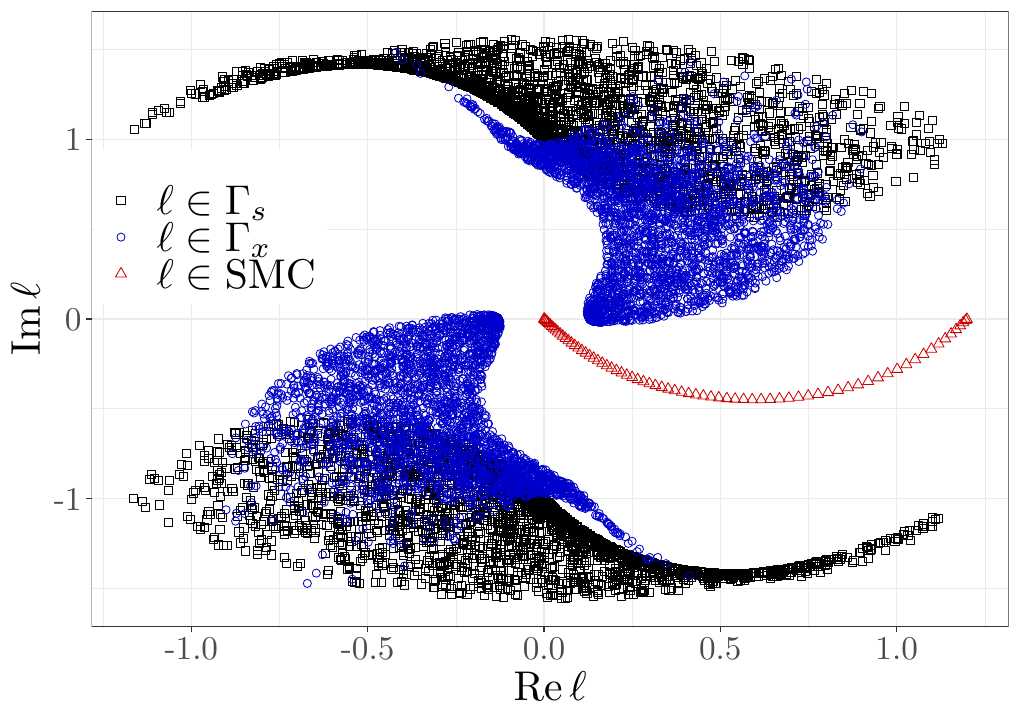} 
    \includegraphics[width=0.49\linewidth]{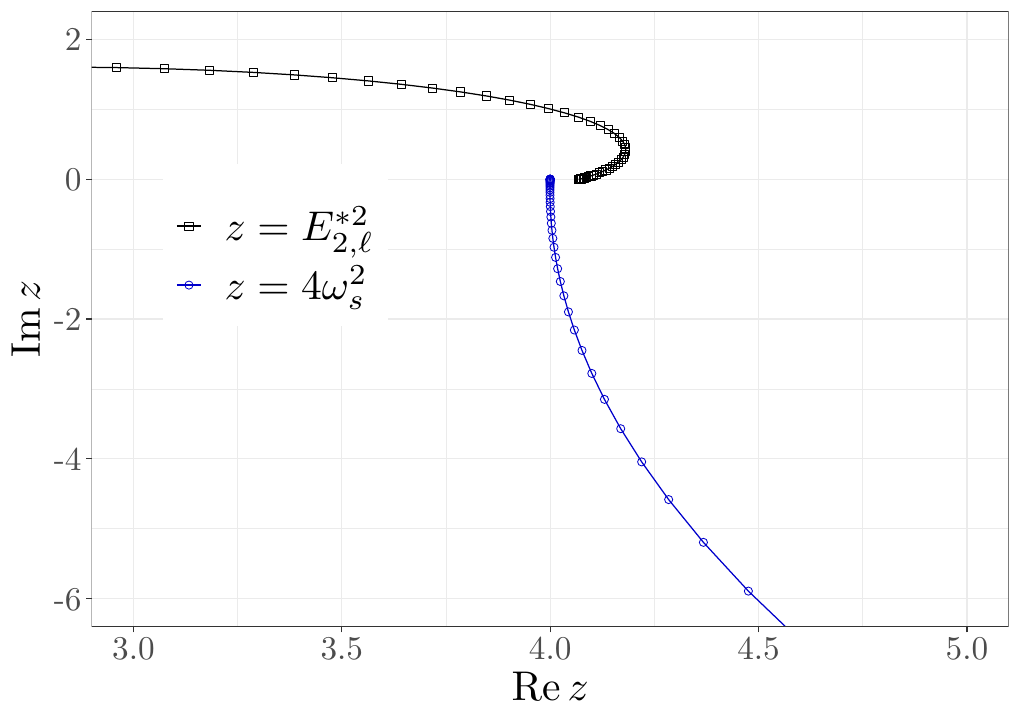}
    \caption{
    On the left panel the integration contour $\ell\in {\rm SMC}$ is shown. It is chosen to avoid the regions in the complex plane $\Gamma_s=\{\ell|\omega_{\ell+k}= 0,\forall_{k\in {\rm SMC}}\}$ and $\Gamma_x=\{\ell|E_3-\omega_{\ell}-\omega_{k}-\omega_{\ell+k}= 0,\forall_{k\in {\rm SMC}}\}$ for a representative three-body energy $E_3/M_0=3.017-0.001i$.
    %$\Delta_1=\omega_{k+p}$ and $\Delta_2=E_3-\omega_{k+p}-\omega_k-\omega_p$ are presented, when both $k$ and $p$ run along the SMC for a representative value for the three-body energy $E_3/M_0=3.017-0.001i$. For completeness, the quantity $\ell$ running along the SMC is shown on the same plot. 
    In the right panel, the quantities $E_{2,\ell}^{*\,2}$ and $4\omega_s^2$ are displayed, where $\ell$ and $s$ run along the SMC and SEC, respectively, for the same value of $E_3$.%, see the text for more details.
    %Representation of the SMC (left panel) and SEC contours (right panel) utilized in this work, cf Eq.~\eqref{eq:SMC/SEC}. For the SMC the .
    }
    \label{fig:FVU:contour}
\end{figure*}
%%%%%%%%%%%%%%%%%%%%%%%%
%%%%%%%%%%%%%%%%%%%%%%%%

%%%%%%%%%%%%%%%%%%%%%%%%
\subsection{Pole position in the FVU approach}
\label{sec:IVU}
%%%%%%%%%%%%%%%%%%%%%%%%

In the FVU approach, the infinite-volume scattering amplitude is extracted as follows. First, the two-body scattering amplitude is simply proportional to $1/(\Sigma^{IV}-\tilde K^{-1}_2)$, whereas the three-body analog is more complex. In particular, the connected isobar-spectator scattering amplitude projected to the $S$-wave reads
%%%%%%%%%%%%%
\begin{align}
\begin{split}
\label{eq:FVU:infvol}
    T_{00}(k,p;E_3)=B_{00}&(k,p;E_3)+C(E_3)\\&-\int_0^\infty \frac{d\ell\,\ell^2}{4\pi^2\omega_\ell}\left(B_{00}(k,\ell;E_3)+C(E_3)\right)\tau(E_{2,\ell}^*)T_{00}(\ell,p;E_3)\,,
\end{split}
\end{align}
%%%%%%%%%%%%%
where $\ell=|\bm{\ell}|$ and $1/\tau(E_{2,\ell}^*)=-q^*_{2,\ell}\cot\delta/(16\pi E_{2,\ell}^*)-i \,{\rm Im\,\Sigma^{IV}(E_{2,\ell}^{*2})}$, see Eq.~\eqref{eq:FVU:SigmaIV}. Here, the $S$-wave projected one-particle exchange is calculated from Eq.~\eqref{eq:FVU:B} as
%%%%%%%%%%%%%
\begin{align}
\label{eq:FVU:B00}
    &B_{00}(k,p;E_3)=\frac{1}{4\pi}\int d\Omega_{\hat p}d\Omega_{\hat k} Y_{00}^*(\hat k)B(\bm{k},\bm{p};E_3)Y_{00}(\hat p)\,.
\end{align}
%%%%%%%%%%%%%
The connection between $T_{00}$ and the three-body scattering amplitude ${\cal M}_3$ can be found in Ref.~\cite{Mai:2017vot}, here we ony need that the poles of 
$T_{00}$ are the same of ${\cal M}_3$.
The complexity in solving the one-dimensional integral equation~\eqref{eq:FVU:infvol} lies in the fact that the interaction kernel (one-particle exchange term $B_{00}$) develops non-trivial cuts. Here, we use the method of the integration contour deformation~\cite{Hetherington:1965zza}, see also Refs.~\cite{Doring:2009yv,Sadasivan:2021emk} for recent applications. One begins with choosing a complex spectator momentum contour (SMC), along which the integration in $\ell$ is performed. The choice is made to ensure one does not hit the singularities of the kernel $B$, Eq.~\eqref{eq:FVU:B}, i.e., zeroes of $\omega_{\ell+k}(E_3-\omega_{\ell+k}-\omega_{\ell}-\omega_{k})$. In practice, this is an iterative process, since the momenta $\ell$ and $k$ should be also located on this contour. These in turn determine the values of $\omega_{\ell+k}$ and $E_3-\omega_{\ell+k}-\omega_{\ell}-\omega_{k}$. The latter should not become zero when both $\ell$ and $k$ are taken somewhere on the contour (otherwise, one would have to choose another contour). The blue and black points in
the left panel of Fig.~\ref{fig:FVU:contour} demonstrate this explicitly. They are generated as follows. The real part of the quantity $E_3$ is fixed somewhere near the expected location of the resonance (We draw the figure for $E_3/M_0=3.017-0.001i$, but we have convinced ourselves that the picture remains the same in the relevant interval of the values of $E_3\in\mathds{C}$). 
Then, the different values of momenta $k$ are chosen on the contour and equations $\omega_{\ell+k}=0$ and $E_3-\omega_{\ell+k}-\omega_{\ell}-\omega_{k}=0$ are solved for $\ell$, which defines two regions, denoted by blue and black dots in the figure. It is seen that the blue and the black areas do not cross the red contour $\ell \in \mbox{SMC}$ and thus the denominator never vanishes. Hence, the singularity of the kernel $B$ is indeed avoided.

In the next step, one picks the  self-energy momentum contour (SEC), along which the integration over the momentum $s$ is performed in $\Sigma^{IV}$, see Eq.~\eqref{eq:FVU:SigmaIV}. This should be done in order to ensure that the integrand in this integral never becomes singular. The right panel of Fig.~\ref{fig:FVU:contour} shows the quantities $E^{*2}_{2,\ell}$ and $4\omega_s^2$ where $\ell$ and $s$ run along the SMC and SEC, respectively, and the same value for $E_3$ is chosen. It is seen that these two quantities never coincide and, hence, the integrand always stays regular.

The choice of the contours is, in principle, a matter of taste. In the present work, we have adopted the following choice:
\begin{align}
{\rm SMC:}&~~\{t-i\,0.6(1-e^{-t/0.3})(1-e^{(t-\Lambda)/0.3})|~t\in(0,\Lambda)\}\nonumber\\
{\rm SEC:}&~~\{t-i\,1.675\arctan\left(0.6t\right)|~t\in(0,\infty)\}\,,
\label{eq:SMC/SEC}
\end{align}
which does not hit any singularities for the range of energies considered in this work. 
In Fig.~\ref{fig:FVU:contour} we have chosen the hard cutoff at $\Lambda=1.2M_0$, see section~\ref{sec:FVU}. For the self-energy integration in Eq.~\eqref{eq:FVU:SigmaIV} the cutoff can be safely removed, since the over-subtracted integrand falls sufficiently quickly at large integration momenta.

Transforming now the (regular) integrals along these contours into finite sums, the integral equation~\eqref{eq:FVU:infvol} can be simply solved as a matrix equation
%%%%%%%%%%%%%
\begin{align}
    {\bf T}_{00}(E_3)=
    \frac{1}{{\bf 1}+ 
    \left({\bf B}_{00}(E_3)+{\bf C}(E_3)\right)
    \cdot\bm{W}\cdot{\bm \tau}(E_{2,\ell}^*)}
    \cdot
    \left({\bf B}_{00}(E_3)+{\bf C}(E_3)\right)\,,
\label{eq:FVU:T00}
\end{align}
%%%%%%%%%%%%%
where the bold symbols denote matrices over spectator momentum $\ell\in{\rm SMC}$. The integration weights $\mu_\ell$ on the chosen contour  are encoded in the matrix $\bm{W}_{pq}=\delta_{pq}p^2/(4\pi^2\omega_p)\mu_p$. Finally, the resonance poles can be found as roots of the equation
%%%%%%%%%%%%%
\begin{align}
\det[{\bf 1}+ \left({\bf B}_{00}(E_3)+{\bf C}(E_3)\right)
    \cdot\bm{W}
    \cdot{\bm \tau}(E_{2,\ell}^*)]=0\,.
\label{eq:FVU:det-pols}
\end{align}
%%%%%%%%%%%%%
Owing to the fact that the integration contour lies in the lower half of the complex energy plane, the quantity ${\bf T}_{00}(E_3)$ is automatically evaluated on the second Riemann sheet for $\Im E_3<0$. We refer the reader to the  Refs.~\cite{Doring:2009yv,Sadasivan:2021emk} for more details.

%%%%%%%%%%%%%%%%%%%%%%%%%%%%%%%%%%%%%%%
\subsection{Pole position in the RFT approach}
%%%%%%%%%%%%%%%%%%%%%%%%%%%%%%%%%%%%%%%

In the case of RFT, the divergence-free scattering amplitude is given by:
\begin{equation}
{\cal M}_\text{df,3}( k_i; p_i) = 
{\cal M}_3( k_i; p_i) - {\cal S}\left\{\mathcal D^{(u,u)}(\bm k, \bm p)\right\}
\,,
\label{eq:Mdf}
\end{equation}
where ${\cal M}_3$ is the full scattering amplitude that depends on the four-momenta of incoming and outgoing particles, $\mathcal D^{(u,u)}$ is a subtraction term that cancels physical divergences present in three-particle scattering, $\cal S$ is a symmetrization operator that sums over the three choices of spectator momentum for both initial and final state, and $\bm k$ and
$\bm p$ are the spectator momenta. Since we focus on $S$-wave interactions, we omit partial-wave indices in the interacting pair.

A resonance appears as a pole in $\mathcal M_3$ in the complex plane, which is inherited by ${\cal M}_\text{df,3}$. Note that the pole position in $\mathcal M_3$ is the same as in the quantity $T_{00}$ from \ref{eq:FVU:T00}.
Explicitly, in the isotropic approximation, ${\cal M}_\text{df,3}$ is given by:
\begin{align}
\Mdf (E_3^*) &= 
\cS\left\{\cL(\bm k) \frac1{1/\Kiso + F_3^\infty} \cR(\bm p) \right\}\,.
\label{eq:M3dKiso}
\end{align}
The quantities $\cL(\bm k)$, $\cR(\bm p)$ and $F_3^\infty$ will be defined below. 
Since the numerator is not divergent, it suffices
to find complex roots of $1/\Kiso + F_3^\infty = 0$. Moreover, in the isotropic limit, all involved quantities are only functions of the energy. All necessary equations to evaluate these quantities are  given here:
\begin{align}
\begin{split}
&\cD_s^{(u,u)}( p,  k) = - \cM_2^s(E_{2,p}^*) G_s( p,  k, \epsilon) \cM_2^s(E_{2,k}^*) \\
&\quad\quad\quad\quad\quad\quad\quad\,- \cM_2^s(E_{2,p}^*) \int_{0}^{k_\text{max}} \frac{k'^2dk'}{(2\pi)^2\omega_{ k'}}  G_s(p,k', \epsilon) \cD^{(u,u)}_s(k', k)
\,,  \\
 & G_s(p,k, \epsilon)
    =-\frac{H(p)H(k)}{4pk} 
    \log\left[ \frac{2pk-(E_3-\omega_{ k}-\omega_{ p})^2+p^2+k^2+m^2-i\epsilon}
    {-2pk-(E_3-\omega_{ k}-\omega_{ p})^2+p^2+k^2+m^2-i\epsilon}\right]\,.
%     \\
% &\frac1{\cM_2^s(k)} =   \frac{ q_{2,k}^* \cot \delta}{16 \pi E_{2,k}^*} + \rho( k)
% %\frac1{\cM_2^s(k)} &=   \frac{ k \cot \delta}{16 \pi E_{2,k}^*} + \rho( k)
% \,,
% \quad 
% \rho(  k) =  \frac{1}{16 \pi E_{2,k}^*}
% \begin{cases}
% -i q_{2,k}^* & E_{2,k}^{* 2} \ge 4 m^2 \,;
% \\
% |q_{2,k}^*| & E_{2,k}^{* 2} < 4 m^2 \,.
% \end{cases} \\
% &\cR( k)  = \cL( k) = \tfrac13 - 2 \omega_{ k} \cM_2^s( k) \wt \rho(k) 
% - \int_{0}^{k_\text{max}} \frac{k'^2dk'}{(2\pi)^2\omega_{ k'}} \cD^{(u,u)}_s( k, k')
% \wt \rho(k')\,, \\
% &F_3^\infty = \int \frac{k^2dk}{(2\pi)^2} \wt \rho( k) \cL( k)\,, \quad \wt \rho( k) = \frac{H( k) \rho( k)}{2\omega_{  k}}
\label{eq:inteqsRFT}
\end{split}
\end{align}

In the formulae above, $k$ and $p$ are the magnitudes of the 
three-momenta while $\epsilon$ is a positive parameter necessary to define $G_s$. Note that we have been working with a finite $\epsilon$, in order to avoid the singularities on the real axis, and the physical solution can be obtained by taking the $\epsilon\to 0$ limit of the subsequent solutions. The cutoff function $H$ is defined in (\ref{eq:Hdef}) while $\cM_2^s(k)$ is the physical s-wave
two-particle scattering amplitude
\begin{align}
&\frac1{\cM_2^s(k)} =   \frac{ q_{2,k}^* \cot \delta}{16 \pi E_{2,k}^*} + \rho( k)
%\frac1{\cM_2^s(k)} &=   \frac{ k \cot \delta}{16 \pi E_{2,k}^*} + \rho( k)
\,,
\quad 
\rho(  k) =  \frac{1}{16 \pi E_{2,k}^*}
\begin{cases}
-i q_{2,k}^* & E_{2,k}^{* 2} \ge 4 m^2 \,;
\\
|q_{2,k}^*| & E_{2,k}^{* 2} < 4 m^2 \,.
\end{cases} 
\end{align}
Finally the quantities $\cL(\bm k)$, $\cR(\bm p)$ and $F_3^\infty$
\begin{align}
&\cR( k)  = \cL( k) = \tfrac13 - 2 \omega_{ k} \cM_2^s( k) \wt \rho(k) 
- \int_{0}^{k_\text{max}} \frac{k'^2dk'}{(2\pi)^2\omega_{ k'}} \cD^{(u,u)}_s( k, k')
\wt \rho(k')\,, \\
&F_3^\infty = \int \frac{k^2dk}{(2\pi)^2} \wt \rho( k) \cL( k)\,, \quad \wt \rho( k) = \frac{H( k) \rho( k)}{2\omega_{  k}}.
\end{align}
To solve numerically the expressions in Eq.~(\ref{eq:inteqsRFT}) on the real axis, we follow the procedure outlined in Refs.~\cite{Jackura:2020bsk,Hansen:2020otl}. We namely replace $\int_{0}^{k_\text{max}}dk'$, with a discrete sum $\sum_{k'}\Delta k$ containing $N$ terms. Then, the first expression in Eq.~(\ref{eq:inteqsRFT}) becomes
\begin{equation}
   {\bf D}(N,\epsilon)={\bf M}\cdot {\bf G}(\epsilon) \cdot{\bf M} - {\bf M}\cdot {\bf G}(\epsilon)\cdot {\bf P} \cdot {\bf D}(N,\epsilon) 
\end{equation}
with the $N\times N$ matrices
\begin{gather}
    {\bf G}_{p,k}(\epsilon)=G_s(p,k,\epsilon)\,,\quad 
    {\bf M}_{p,k}=\delta_{pk}\cM_2^s(E_{2,p}^*)\,,\\
    {\bf P}_{p,k}=\delta_{pk}\frac{k^2\Delta k}{(2\pi)^2 \omega_{ k}}
\end{gather}
and 
\begin{gather}\label{eq:lim-Duu}
    \cD^{(u,u)}_s( p,  k) =\lim_{N\to\infty}\lim_{\epsilon\to 0}{\bf D}(N,\epsilon)\,.
\end{gather}
Once this function is available, everything else is straightforward to evaluate. 

\begin{figure}[th]
    \centering
    \includegraphics[scale=0.42]{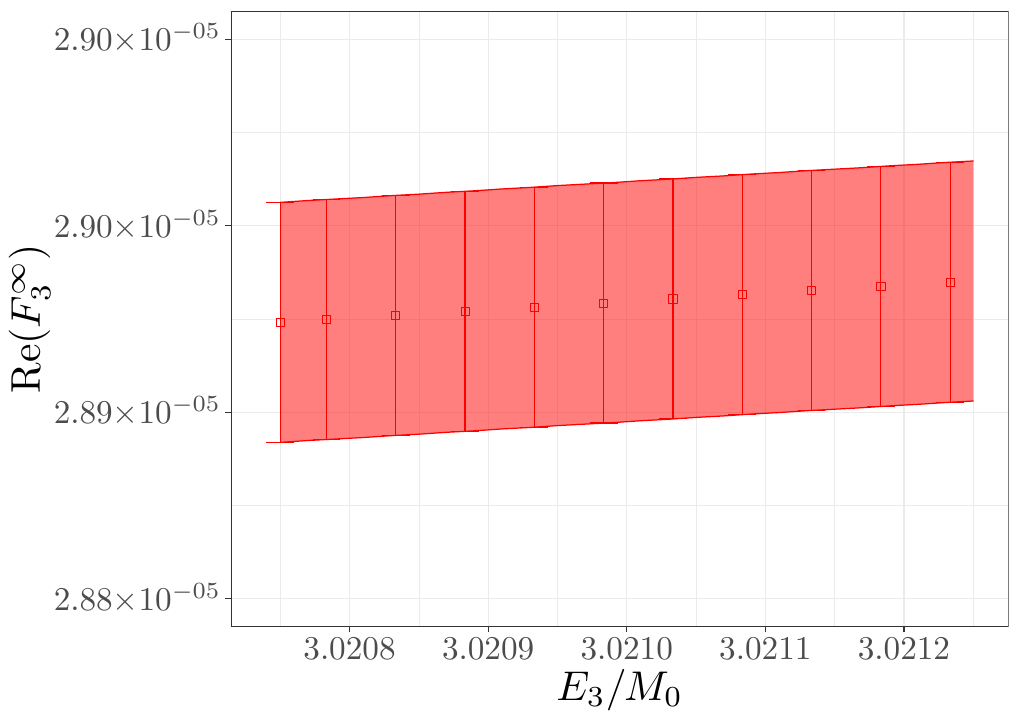}
    \includegraphics[scale=0.42]{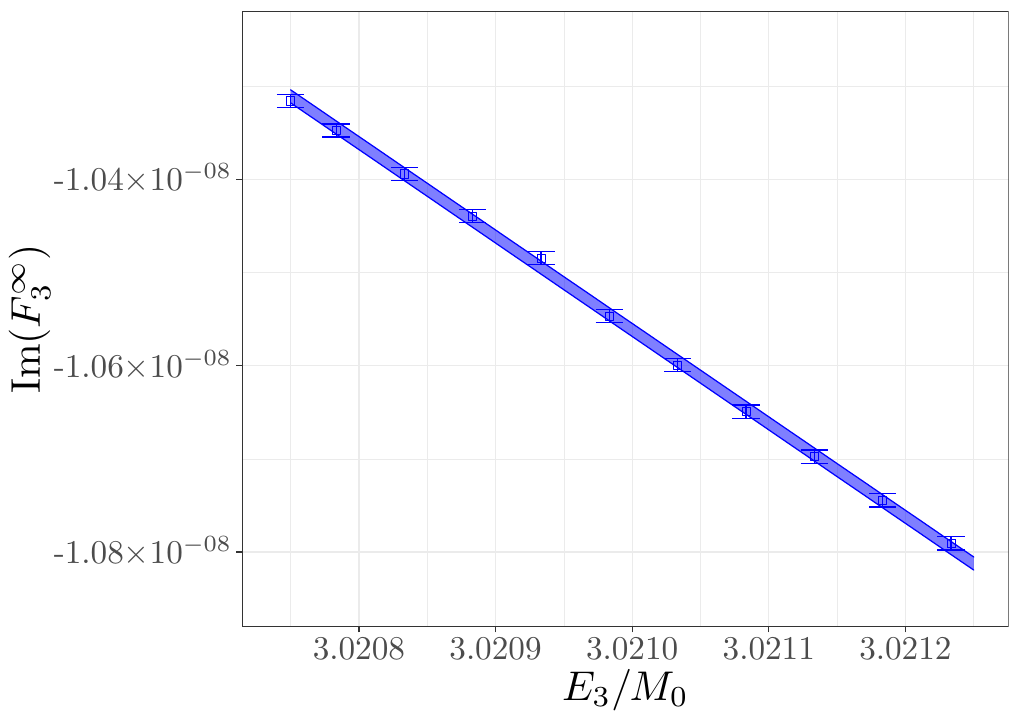}
    \caption{Real and imaginary part of $F_3^\infty$ for $g=17.81$, in the small interval $E_3/M_0\in [3.02075,3.02123]$ a line is sufficient to interpolate the data. }
    \label{fig:Finf_reim}
\end{figure}

The final step is to perform an analytic continuation into the complex plane. 
However, since the interaction is weak we expect the resonance to be very close to the real axis, and we can simply extrapolate from the real axis to the complex plane. This avoids issues with the analytic continuation of the cutoff function in Eq.~\eqref{eq:inteqsRFT}. More specifically, we fit the real and imaginary part of $F_3^\infty$ to a simple polynomial in energy to build an interpolating function, see Fig.~\ref{fig:Finf_reim}. An example of this is shown in Fig.~\ref{fig:Finf_reim}. Finally, we use that function to find the zeros of the denominator of Eq.~\eqref{eq:M3dKiso}. 

%%%%%%%%%%%%%%%%%%%%%%%%%%
\subsection{Results for the mass and the width of the resonance}
\label{subsec:pole_position}
%%%%%%%%%%%%%%%%%%%%%%%%%%

Once the three-body forces are determined as in section~\ref{subsec:comparison_FVU_RFT}, we need to solve an integral equation in both approaches FVU and RFT to extract the physical information. 
We solve the RFT integral equation using  $\epsilon=10^{-7}$ and $N=2000$,  while in FVU we discretize the contour with 200 points,
in both cases we did not observe any residual discretization effects.
% where we do not observe any residual effects from the limit in Eq.~(\ref{eq:lim-Duu}).
The final step is to find positions in the complex plane, such that the three-to-three amplitude has a pole:
%%%%%%%%%%%%%%%%%%%%%%%%%%%%%%%%%%%%%%%
\begin{equation}
    {\cal M}_3=\frac{R_{-1}}{E_3-M_R+i\Gamma/2}+R_0\,,
\end{equation}
%%%%%%%%%%%%%%%%%%%%%%%%%%%%%%%%%%%%%%%
for a range of energy $|E_3-M_R|<\Gamma$, the above is known
as Breit-Wigner parametrization. In the FVU, the pole positions are extracted directly on the second Riemann sheet, calculating then the mass and the width of the resonance via $M_R-i\Gamma/2=E_3^*$.

\begin{figure}[t]
    \centering
    \includegraphics[width=0.85\linewidth]{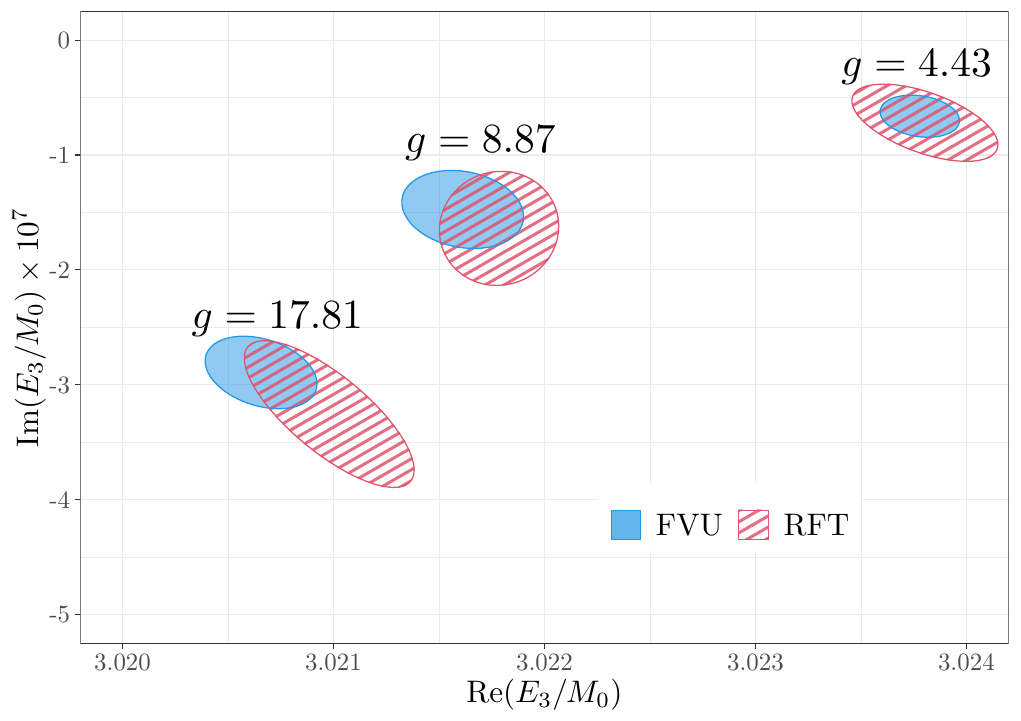}
    \caption{Comparison of the pole positions between the FVU and RFT three-particle formalisms. The pole position is related to the mass and width of the resonance reported in Table~\ref{tab:pole_M3}.}%[9/8/22]
    \label{fig:FVU_vs_QFT_poles1}
\end{figure}

\begin{figure}[t]
    \centering
    \includegraphics[width=0.85\linewidth]{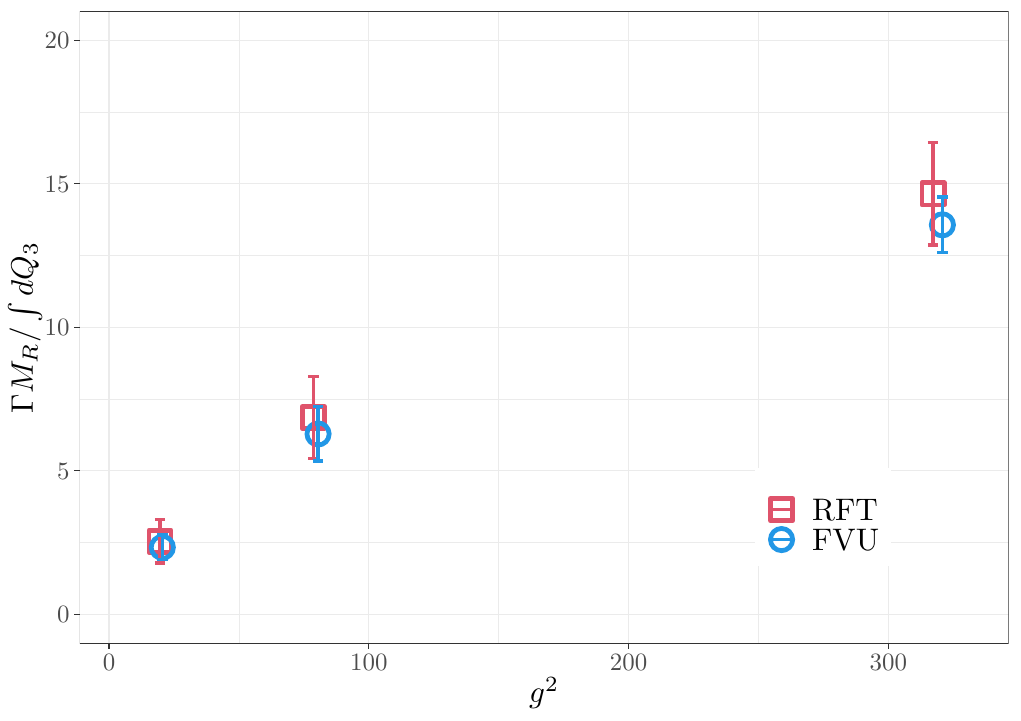}
    \caption{Values of $\Gamma$ multiplied by $M_R$ normalized with the three particle phase space $\int d Q_{1\to3}$ computed in RFT (red squares) and FVU (blue circles). }
    \label{fig:Gamma_vs_g}
\end{figure}

In Fig.~\ref{fig:FVU_vs_QFT_poles1} we observe that FVU and RFT give compatible predictions of the pole position within errors and, thus, physical parameters $M_R$ and $\Gamma$ as reported in Table~\ref{tab:pole_M3}.
As expected, the width increases with increasing values of the coupling $g$.
The decay width of one particle into three identical particles can be computed as 
\begin{equation}
    \Gamma = \frac{1}{2M_R 3!}\int dQ_{1\to3} |{\cal M}_{1\to3}|^2\,,
    \label{eq:Gamma_LO}
\end{equation}
where the factor $1/3!$ is a symmetry factor taking into account that the final particles are identical while $\int dQ_{1\to3}$ is the integral over the three-particle phase space which for total momentum $\bm P=0$ reads
\begin{equation}
    dQ_{1\to3}=(2\pi)^4\delta^4(p_1+p_2+p_3-P)\prod_{i=0}^3\frac{d\bm{p}_i}{(2\pi)^32 \omega_{{p}_i}}\,,
\end{equation}
and ${\cal M}_{1\to3}$ is the scattering amplitude with one initial $\phi_1$ and three final $\phi_0$.
As a matter of fact, the phase space factor is responsible for the small size of the width of the found resonance. To exemplify this and also to compare the obtained widths with the tree level expectation ($\sim g^2$), we plot the ratio $\Gamma M_R/\int dQ_{1\to3}$ as a function of $g^2$ in Fig.~\ref{fig:Gamma_vs_g}. We observe that the results for the combination $\Gamma M_R/\int dQ_3$ have a slope of ${\cal O}(10^{-1})$. Furthermore, for lower values of $g$, the relation seems linear but lower than the tree-level prediction $O(1)$\footnote{Here, it is assumed that a bare value of coupling $g$ can be used to obtain numerical predictions}. Finally we observe deviation from the linearity for the highest point in $g$.
% non-linear effects appear to be non-negligible at higher $g$.

%%%%%%%%%%%%%%%%%%%%%%%%
\begin{table}[t]
    \centering
    \begin{tabular}{cccc}
    \hline
         $g$ &    & $M_R/M_0$ & $\Gamma/M_0\times10^{7}$  \\
         \hline
         4.43 & FVU  &   3.02378 (12) &  1.32 (24) \\
         4.43 & RFT  &   3.02383 (22) &  1.45 (43) \\
         \hline
         8.87 & FVU  &   3.02161 (19) &  2.95 (45) \\ 
         8.87 & RFT  &   3.02179 (19) &  3.27 (68) \\
         \hline
         17.81 & FVU  &   3.02066 (17) &  5.79 (41) \\
         17.81 & RFT  &   3.02098 (26) &  6.48 (79) \\
    \hline
    \end{tabular}
    \caption{Values of the mass and width of the resonance $E_3^\text{pole}=M_R-i\Gamma/2$ 
    computed in FVU and RFT using the parametrizations of the three-body force of Eq.~(\ref{eq:3bf-FVU/RFT}) and the best fits from Table~\ref{TAB:2&3body-fits}. The error reported is only statistical.}
    \label{tab:pole_M3}
\end{table}
%%%%%%%%%%%%%%%%%%%%%%%%

% \pagebreak
%%%%%%%%%%%%%%%%%%%%%%%%%%%%%%%%%%%%%%%%%%%%%%%%%%
\section{Conclusion}
\label{sec:conclu}
%%%%%%%%%%%%%%%%%%%%%%%%%%%%%%%%%%%%%%%%%%%%%%%%%%

We have determined the properties of resonances with the three-particle decay modes in the complex $\varphi^4$ theory. This has been achieved after several steps:
(i) generating field configurations and computing the finite-volume energy levels, (ii) analyzing the spectrum with (different) finite-volume formalisms, and (iii) solving the integral equations to compute the pole position of the three-particle amplitude in the complex energy plane.

The model of choice contains two complex scalars with masses $M_1>3 M_0$, and an explicit term in the Lagrangian, allowing a one-to-three decay. By solving the Generalized Eigenvalue problem, we have determined the energy levels of two and three particles. Given the affordable computational cost of this theory, we have carried out the simulations at several lattice volumes and parameters in the action. More details about the theory can be found in section~\ref{sec:phi4model}, and a summary of the energy levels is provided in Table~\ref{tab:latt-spectrum}. 

Regarding the analysis of the spectra, we have used two versions of the three-particle finite-volume formalisms: the RFT and FVU. Indeed, this is the first time that the same dataset has been analyzed using the two formalisms.  By fitting the energy levels using the quantization conditions, we have obtained the two- and three-body scattering parameters. Our findings support the statement that comparable descriptions of the finite-volume spectrum can be achieved with either formalism, i.e., with similar $\chi^2$ in the fits. Figure~\ref{fig:FVU_vs_QFT} shows the lattice spectra and the different fits with the two approaches. We have found that in order to describe the energy levels and the observed avoided level crossing, an explicit pole in the three-body forces, $\Kdf$ for RFT, and $C$ for FVU, is needed.

The scattering parameters in the three-body sector, obtained with the two formalisms are neither directly comparable nor physical, as they come with a particular scheme- and cutoff dependence. For this reason, we have evaluated physical observables, such as the mass and the width of the resonance. The computation of this quantity involves solving integral equations and performing an analytic continuation into the complex energy plane. In this way, we find completely consistent numerical results for these observables, see Fig.~\ref{fig:FVU_vs_QFT_poles1} for the main result of this work. We can indeed conclude that the physical observables, computed in this work, have a small systematic dependence on the underlying choice of parametrization for the three-body interactions.

We have therefore demonstrated the practical equivalence of the different available three-body methods in this controlled setup.  Future work will  involve applying the same steps to the QCD resonances. Some additional complications will then be needed to be addressed (e.g. nonidentical particles, multichannel scattering, spin. etc.), but the workflow presented in this work will generally remain.

%%%%%%%%%%%%%%%%%%%%%%%%%%%%%%%%%%%%%%%%%%%%%%%%%%
\acknowledgments
%%%%%%%%%%%%%%%%%%%%%%%%%%%%%%%%%%%%%%%%%%%%%%%%%%
We thank C. Lang, M. Döring and D. Sadasivan for useful discussions. This work is supported by the Deutsche Forschungsgemeinschaft (DFG, German Research Foundation), the NSFC through the funds provided to the Sino-German Collaborative Research Center CRC 110 “Symmetries and the Emergence of Structure in QCD” (DFG Project-ID 196253076 -
TRR 110, NSFC Grant No.~12070131001) and by the Ministry of Culture and science of North Rhine-Westphalia thought the NRW-FAIR project. MM was further supported by the National Science Foundation under Grant No. PHY-2012289. 
FRL has been supported in part by the U.S. 
Department of Energy (DOE), Office of Science, Office of Nuclear Physics, under grant Contract Numbers 
DE-SC0011090 and DE-SC0021006.  In addition, the work of AR was funded in part by the Volkswagenstiftung (grant no. 93562) and the Chinese Academy of Sciences (CAS) President’s International Fellowship Initiative (PIFI) (grant no. 2021VMB0007).

%%%%%%%%%%%%%%%%%%%%%%%%%%%%%%%%%%%%%%%%%%%%%%%%%%

%%%%%%%%%%%%%%%%%%%%%%%%%%%%%%%%%%%%%%%%%%%%%%%%%%
\bibliographystyle{JHEP} 
\bibliography{bibliography}
%%%%%%%%%%%%%%%%%%%%%%%%%%%%%%%%%%%%%%%%%%%%%%%%%%

\end{document}